\documentclass[10pt]{iopart}
\usepackage{iopams}  
\usepackage{xcolor}
\usepackage{graphicx}
\usepackage{hyperref}
\hypersetup{
    colorlinks=true,
    linkcolor=blue,
    filecolor=magenta,      
    urlcolor=cyan,
    pdftitle={Study of cognitive component of auditory attention to natural speech events},
    pdfpagemode=FullScreen,
}
\usepackage{subcaption}
\usepackage{iopams}
\usepackage{amsmath}
\usepackage{amssymb}
\usepackage{mathtools}
\usepackage{multirow}

\DeclareMathOperator*{\argmax}{argmax}
\newcommand\norm[1]{\left\lVert#1\right\rVert}

\begin{document}

\title[Study of cognitive component of auditory attention to natural speech events]{Study of cognitive component of auditory attention to natural speech events}

\author{Nhan D T Nguyen, Kaare Mikkelsen and Preben Kidmose}

\address{Department of Electrical and Computer Engineering, Aarhus University, Aarhus, Denmark}
\ead{ndtn@ece.au.dk}
\vspace{10pt}
\begin{indented}
\item September 2023
\end{indented}

\begin{abstract}

\noindent{\it Objective.} Event-related potentials (ERP) have been used to address a wide range of research questions in neuroscience and cognitive psychology including selective auditory attention. The recent progress in auditory attention decoding (AAD) methods is based on algorithms that find a relation between the audio envelope and the neurophysiological response. The most popular approach is based on the reconstruction of the audio envelope based on EEG signals. However, these methods are mainly based on the neurophysiological entrainment to physical attributes of the sensory stimulus and are generally limited by a long detection window. This study proposes a novel approach to auditory attention decoding by looking at higher-level cognitive responses to natural speech. 
{\it Approach.}
To investigate if natural speech events elicit cognitive ERP components and how these components are affected by attention mechanisms, we designed a series of four experimental paradigms with increasing complexity: a word category oddball paradigm, a word category oddball paradigm with competing speakers, and competing speech streams with and without specific targets. We recorded the electroencephalogram (EEG) from 32 scalp electrodes and 12 in-ear electrodes (ear-EEG) from 24 participants.
{\it Main results.}
 By using natural speech events and cognitive tasks, a cognitive ERP component, which we believe is related to the well-known P3b component, was observed at parietal electrode sites with a latency of approximately 620 ms. The component is statistically most significant for the simplest paradigm and gradually decreases in strength with increasing complexity of the paradigm. We also show that the component can be observed in the in-ear EEG signals by using spatial filtering.
{\it Significance.}
The cognitive component elicited by auditory attention may contribute to decoding auditory attention from electrophysiological recordings. Furthermore, the presence of this component in the ear-EEG signals is promising for future applications within hearing aids.
\end{abstract}

%
\vspace{2pc}

%
\maketitle
%
\ioptwocol

\section{Introduction}\label{s_introduction}

The event-related potential (ERP) technique is a non-invasive neurophysiological technique that measures potential fluctuations in the electroencephalogram (EEG) in response to sensory, cognitive or motor events. One of the advantages of ERPs is their high temporal resolution which allows us to measure neural activity with millisecond precision \cite{woodmanBriefIntroductionUse2010}. This provides insights into the timing of cognitive processes that take place right after the delivery of sensory information to the peripheral nervous system until even after a behavioral response is made. Therefore, the ERP method has been widely used to answer different research questions relating to cognitive processes, including perception, attention, memory, language, and decision-making.

A remarkable aspect of human perception is our ability to perceptually segregate concurrent auditory objects and selectively attend to these objects. It is an ability we are readily familiar with and utilize in many situations, for example in cocktail party scenarios, to volitionally focus on the speaker(s) of interest. Decoding of this attention process, based on electrophysiological signals, is referred to as auditory attention decoding (AAD).

Since the inception of ERP research, there has been an interest in using the ERP technique for investigating selective attention mechanisms. Extensive research has been dedicated to understanding which components and cognitive systems are influenced by attention. Here, a fundamental question has been whether the attention mechanism is a phenomenon that happens in the early or late stages of neural processing. Some researchers \cite{broadbentPerceptionCommunication1958, treismanSELECTIVEATTENTIONMAN1964, treismanSelectiveAttentionPerception1967} have hypothesized that it is an early selection mechanism when the sensory systems are overloaded by multiple inputs, leading to the selection of input for further processing. One of the well-known experimental paradigms used to address this hypothesis was introduced by Hillyard \etal \cite{hillyardElectricalSignsSelective1973}. Two sequences of brief tones were presented simultaneously to subjects, with tones at one frequency presented to the left ear and tones at another frequency presented to the right ear. Subjects were instructed to attend to the stimuli presented to one of the ears while ignoring the other one and press a button if they detected a softer/stronger tone (deviant) in the attended sequence. They found that the N1 component (occurring around 150 ms) was larger for the tones presented in the attended ear than those in the unattended ear. Additionally, the N1 effect was the same for both deviant and standard stimuli, indicating that the attention mechanism affects the neural processing in an early stage before the stimulus identification is completed. This supports the early selection hypothesis of attention.

Initially, ERP research predominantly considered transient ERP's, while responses to continuous stimuli like Auditory Steady State Response (ASSR) were introduced later on \cite{galambos40HzAuditoryPotential1981}. Gradually, research moved in the direction of more naturally occurring stimuli, e.g. investigating the entrainment to the temporal envelope of speech signals \cite{aikenHumanCorticalResponses2008}. The previously mentioned segregation mechanism, that helps us in cocktail party situations, is revealed in the electrophysiological brain response as an entrainment to the audio envelope of the auditory objects, where the entrainment is stronger for the attended object \cite{dingEmergenceNeuralEncoding2012}. Thereby, the research in attention mechanisms became relevant for the research in envelope-following responses.

The impact of attention on the envelope-following responses forms the foundation for research in the field of auditory attention decoding. In AAD, the objective is to decode the auditory attention based on the electrophysiological response. The attended speech stream is determined as the one that has the actual envelope most correlating with the reconstructed envelope. Previous studies have used both linear \cite{osullivanAttentionalSelectionCocktail2015, biesmansAuditoryInspiredSpeechEnvelope2017, geirnaertUnsupervisedSelfAdaptiveAuditory2021, aroudiImpactDifferentAcoustic2019, osullivanNeuralDecodingAttentional2017} and non-linear models  \cite{detaillezMachineLearningDecoding2020, nogueiraDecodingSelectiveAttention2019, xuAuditoryAttentionDecoding2022} to estimate the mapping between the audio envelope and the neural response. Another approach, called {\it canonical component analysis (CCA)} \cite{hardleCanonicalCorrelationAnalysis2007}, combines a backward model on the EEG and a forward model on the speech envelope to maximize the correlation of their outputs jointly  \cite{decheveigneDecodingAuditoryBrain2018}. However, these methods require a long decision window to achieve high accuracy. The CCA method has so far been the best method \cite{geirnaertElectroencephalographyBasedAuditoryAttention2021} with a decoding accuracy up to approximately 90\% for a decision window length of 30 seconds tested on specific datasets in a very well-controlled environment, with two competing speakers.

In parallel with the research on early attention mechanisms, another line of research focused on attention mechanisms in the post-perceptual processes \cite{morayAttentionDichoticListening1959, deutschAttentionTheoreticalConsiderations1963}. Luck \etal \cite{luckERPComponentsSelective2011} stated that selective attention could happen in any system when that system becomes overloaded. That means attention may happen at the early or late stage or even both depending on the nature of stimuli and task. For instance, we can design experiments in which subjects must perform several complicated cognitive tasks and respond accordingly. This could lead to an overload in the memory and response systems and in consequence an impact on the late post-perceptual components. Therefore, late cognitive components like the N400 and P300 become relevant in the studies of attention. N400 is a slow negative-going deflection between 200 and 600 ms and is elicited by semantic mismatch. In one study \cite{graingerWatchingWordGo2009}, using a word priming paradigm, it was found that less perceptible priming words resulted in an attenuation of the N400 component. In another study \cite{erlbeckTaskInstructionsModulate2014}, attention-manipulated tasks (active, passive, ignore) were used to investigate the effect of attention on auditory N400 and the Mismatch Negativity (MMN). The authors concluded that attention has an effect on both N400 and MMN components. On the other hand, the P300 component is an endogenous potential, as it is related to cognitive processing rather than the physical attributes of a stimulus. More specifically, the P300 is thought to reflect processes involved in stimulus evaluation or categorization. The P300 has a positive-going amplitude that peaks at around 300 milliseconds, and the peak will vary in latency from 250 to 500 milliseconds or more, depending upon the stimuli, the task, and the subject. Due to the component’s latency, the term P300 was first used to name this component. However, many researchers prefer to use the name {\it P3} to refer to the third positive deflection which almost always peaks well after 300 ms. To avoid confusion, the remaining part of this paper uses the name {\it P3} to refer to this component. The P3 is commonly divided into two sub-components:{\it P3a}, which reflects the novelty of the rare events, and {\it P3b}, which is elicited when relevant tasks involving rare events are required. Numerous studies have investigated how auditory attention affects the P3. Other studies \cite{isrealP300TrackingDifficulty1980, kramerAnalysisProcessingRequirements1983, mangunAllocationVisualAttention1990} showed that the more attention or resources allocated to process stimulus, the larger the P3 amplitude is. This suggests that the cognitive component P3 has the potential to be a reliable feature for decoding auditory attention.

The majority of AAD research relies on electrophysiological measurement methods, such as electroencephalography, magnetoencephalography, and electrocorticography, primarily for research purposes. Despite their primary research-oriented application, the insights gained into auditory attention mechanisms hold substantial value for developers of hearing devices to enhance the audio processing algorithms. However, a strong motivating factor for developing ADD methods has been the potential integration of these techniques into hearing devices. To make this feasible in reality, it necessitates the development of less invasive methods for capturing electrophysiological signals than those known from conventional electrophysiological measurement systems. As a response to this challenge, there has been an increasing interest in the development of miniaturized and wearable EEG devices that provide a discrete, unobtrusive, and user-friendly recording solution. One such solution is so-called ear-EEG systems \cite{kappelDryContactElectrodeEarEEG2019}, where EEG is recorded from electrodes placed in-the-ears. Numerous studies have compared ear-EEG and conventional scalp EEG, e.g. in terms of auditory middle and late latency ERPs, signal-to-noise ratio, power spectrum \cite{mikkelsenEEGRecordedEar2015} and P3 response \cite{farooqEarEEGBasedVisual2015}. The ear-EEG setup has also been used to investigate the effect of the repetition rate of chirp stimuli on ASSR \cite{christensenChirpEvokedAuditorySteadyState2022}, and to perform automatic sleep staging and sleep scoring at almost the same levels as other methods \cite{mikkelsenAutomaticSleepStaging2017, mikkelsenAccurateWholenightSleep2019}. Additionally, in the field of AAD, in-ear single sensor \cite{fiedlerSinglechannelInearEEGDetects2017} and around-the-ear sensors \cite{holtzeEarEEGMeasuresAuditory2022} have successfully been demonstrated to unobtrusively monitor auditory attention to tone streams and continuous speech streams.

Although ADD methods based on envelope-following responses have come a long way, there are still significant limitations e.g. in the decoding accuracy and the length of the decision window. This has motivated us to investigate endogenous components related to auditory attention. Thus, this paper investigates the cognitive components associated with auditory attention using natural language events. The study is based on a series of experiments with speech signals, where the complexity of the paradigms gradually increases, from a single speaker and isolated words to competing speakers and continuous speeches. Additionally, we conduct a comprehensive comparative analysis of both scalp and in-ear EEG data to assess the feasibility of using the studied components to solve the AAD problem.

\section{Methods}\label{s_methods}

\subsection{Participants}\label{ss_participants}
Twenty-four native Danish-speaking subjects (26.87 ± 8.44 years, 13 male, 5 left-handed) participated in the study. All subjects provided written informed consent and reported normal hearing ability and no neural disorders. The experimental protocol was approved by the Research Ethics Committee (Institutional Review Board) at Aarhus University, approval number 2021-79.

\subsection{Experimental paradigms}\label{ss_prdg}
The study comprised four experimental paradigms. The paradigms represent an increasing level of complexity, and all participants conducted all four experiments sequentially from paradigm 1 to 4.

\subsubsection{Paradigm 1: Word category oddball.}\label{prdg_1}

This paradigm was designed to be similar to the conventional oddball paradigm in which subjects were presented with a sequence of two different classes of spoken words: \textit{animal names} and \textit{cardinal numbers}, or \textit{color names} and \textit{cardinal numbers} from a loudspeaker situated one meter in front of the subject. The animal names and color names were predefined as the \textit{target} events, while the cardinal numbers were the \textit{non-target} events. The target and non-target events in this context play similar roles as oddball and standard events in the classical oddball paradigm. However, unlike conventional oddball paradigms, the discrimination between the target and non-target events was not based on the physical attributes of the stimuli but rather on the semantics of the stimuli. The stimuli for each trial were generated by randomly mixing twenty target and non-target events, with the number of target events between 2 and 5 and the first two being non-targets. The number of target events was chosen based on two criteria: 1) to provide enough data for analysis, and 2) not too many since less probable events would produce a larger cognitive response. The distance between two consecutive events was random and uniformly distributed between 0.8 and 1.2 seconds, but the total length of the trial was kept at 20 seconds. In each trial, the subject was asked to pay attention to the target events and passively count them. At the end of each trial, the subject reported the number of target events and received feedback on their accuracy. The counting task and feedback were used to encourage the subject to remain engaged in the task. There were sixteen trials in this paradigm. The target events in the first trials were animal names and the target events in the last eight trials were color names.

\subsubsection{Paradigm 2: Word category with competing speakers.}\label{prdg_2}

Paradigm 2 was an extension of paradigm 1, using similar sequences of twenty discrete spoken words. However, instead of a single stream of words, two competing streams were presented simultaneously by two speakers located at equal distances on either side of the subject, placed 60 degrees to the left and right (see \fref{fig2_experimental_setup}). In each trial, the subject was asked to pay attention to only the target events in one of the streams and completely ignore the other stream. The subject was instructed to passively count the number of target events in the attended stream and report the count after the trial. The target events in the twenty trials were balanced between the two classes of events. Additionally, the attended stream was randomly changed but balanced between the left and right speakers to avoid bias towards a particular listening direction.

\subsubsection{Paradigm 3: Competing speech streams with targets.}\label{prdg_3}

In paradigm 3, the setup was similar to the setup of paradigm 2. The subject was presented with two competing streams from the same two speakers as in paradigm 2. However, in this case, the stimuli in each speaker were not sequences of spoken words but snippets of different stories and each snippet had a duration of approximately 20 seconds. Each trial had one class of words predefined as target words. For instance, a target class could be {\it human names}. In each trial, the subject was asked to pay attention to only one of the two streams and focus on the target words of that stream. At the end of the trial, the subject answered a question about the target words and received feedback. There were 4 classes of target words: animal names, human names, color names, and plant species, distributed over 5 different stories. The story of the attended stream in each trial continued from where it ended in the previous trial. This made the stream easier to follow and attend. There were 20 snippets in total. Each snippet appeared twice in two different trials: one time as the attended stream and one time as the unattended stream. The attended stream was also randomized and balanced between the left and right speakers.

\subsubsection{Paradigm 4: Competing speech streams without targets.}\label{prdg_4}

Paradigm 4 was designed to simulate a real-world scenario of selective listening in a setting with multiple sound sources. Two competing streams from the same loudspeaker setup as utilized in paradigms 2 and 3 were presented to the subject, each stream contained a single-speaker narrative. No particular target words were specified for this paradigm. The subject was instructed to attend to one stream while disregarding the other in each trial. Following each trial, the subject was probed with a question about the content and was provided with feedback. The data collected from this experiment was not employed in any analyses or results presented in this paper, but is merely described here for completeness.

The motivation for this experimental design was to investigate the cognitive processing of speech events. Going from paradigm 1 to 3, the experimental complexity was increased in terms of task difficulty and practicality. In this way, we wished to see how the ERP component is regulated by the distinctiveness of the event and the selective listening mechanism.

\subsection{Stimuli}\label{ss_stimuli}
The stimuli of all paradigms were in Danish and were synthesized using the Google Text-to-Speech tool v2.11.1 \cite{TexttoSpeechAILifelike}. The voice configuration was randomly selected between {\it da-DK-Wavenet-A} (female)  and {\it da-DK-Wavenet-C} (male) to generate each snippet in paradigms 3 and 4. In the end, there were 14 out of 20 male voice snippets in paradigm 3 and 24 out of 40 male voice snippets in paradigm 4. Each voice configuration was selected to generate every single word in paradigms 1 and 2 once. The speed was set at 0.85 and the authenticity was verified by native Danish people to make sure that the pronunciation was as natural and clear as possible. In this process, minor changes were made to the text to increase speech quality. The target words in paradigms were selected to be short ( 1 or 2 syllables) to resemble the oddball events. The details of stimuli generation are as below:
\begin{itemize}

\item Paradigms 1 and 2 shared the same words. The sounds of all discrete Danish words were normalized to have the same root mean square (RMS) amplitude. The normalized sound files were then concatenated to fit the experimental design.

\item The stimuli for paradigms 3 and 4 were generated in a similar way. Text scripts of the stories were first made and then used for audio synthesis using the text-to-speech tool. The audio files were then sliced into different snippets and normalized to have the same RMS amplitude. The text scripts for paradigm 3 were created by native Danish speakers. For experiment 4, the text scripts were sourced from two books (The Hobbit and Northern Lights) and excerpts from Danish radio broadcast news covering politics, society, education, sports, and social networks. The word onset times, to be used in the subsequent ERP analysis, was extracted from the text-to-speech tool.
\end{itemize}

\begin{figure}[ht]
 \centering
 \begin{subfigure}[ht]{\linewidth}
     \centering
     \includegraphics[width=\linewidth]{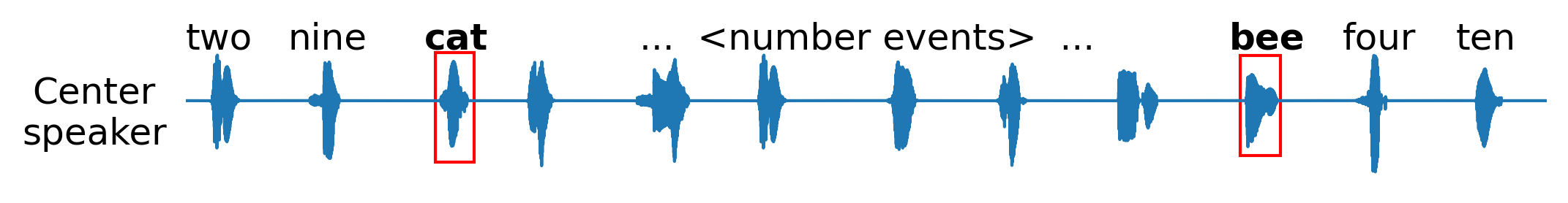}
     \caption{Paradigm 1: Word category oddball.}
     \label{fig1a_stimuli_illustration}
 \end{subfigure}
 \hfill
 \begin{subfigure}[ht]{\linewidth}
     \centering
     \includegraphics[width=\linewidth]{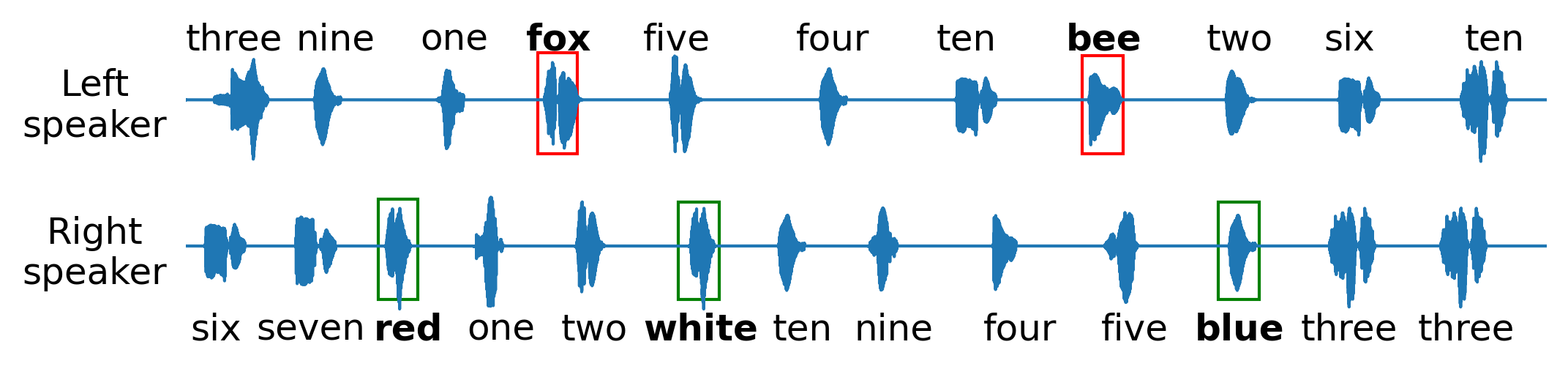}
     \caption{Paradigm 2: Word category with competing speakers}
     \label{fig1b_stimuli_illustration}
 \end{subfigure}
 \hfill
 \begin{subfigure}[ht]{\linewidth}
     \centering
     \includegraphics[width=\linewidth]{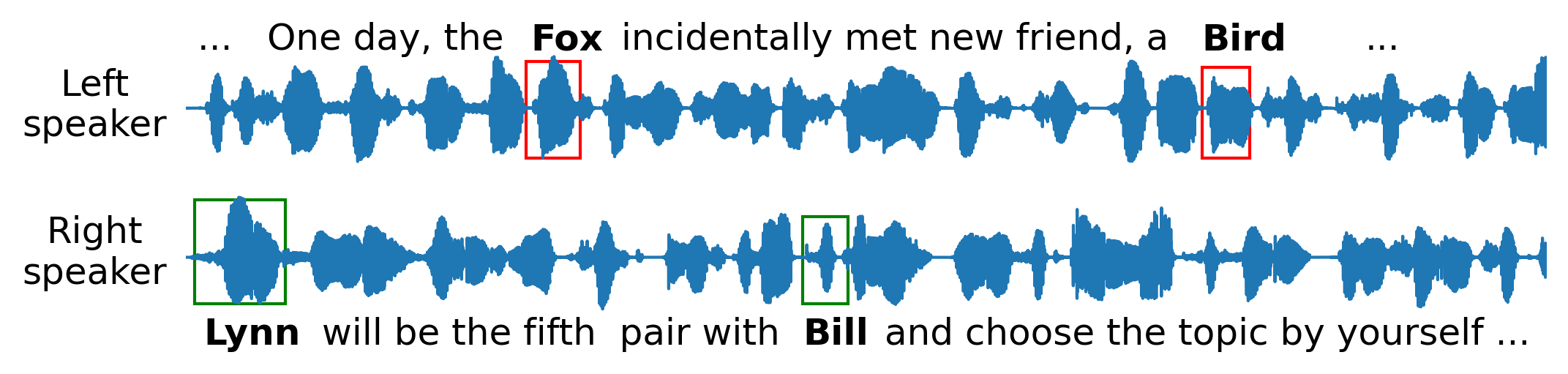}
     \caption{Paradigm 3: Continuous speech with targets.}
     \label{fig1c_stimuli_illustration}
 \end{subfigure}
 \hfill
 \begin{subfigure}[ht]{\linewidth}
     \centering
     \includegraphics[width=\linewidth]{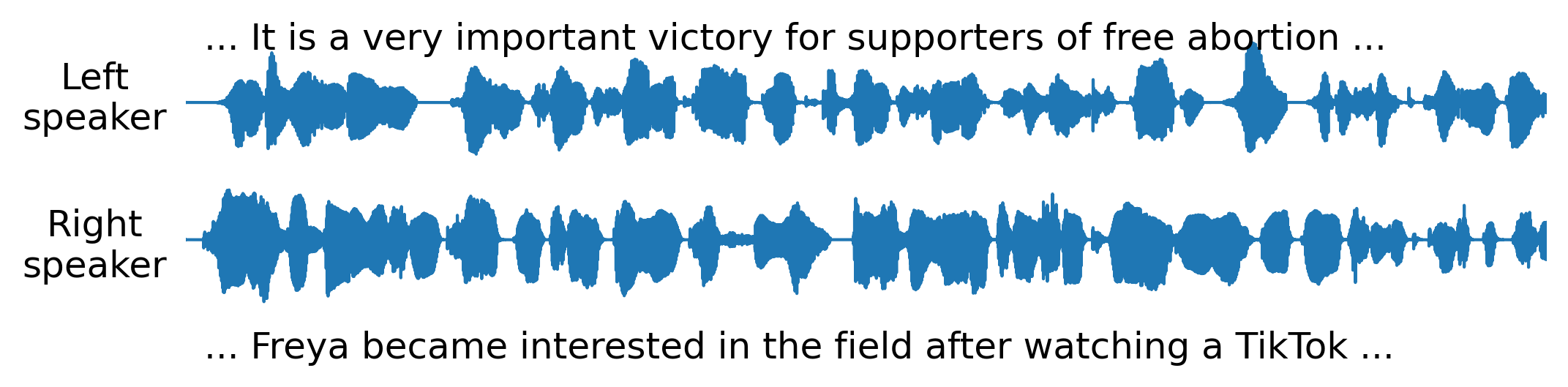}
     \caption{Paradigm 4: Continuous speech without targets.}
     \label{fig1d_stimuli_illustration}
 \end{subfigure} 
\caption{Illustration of the stimuli for the four paradigms. The text above or below the waveforms shows the corresponding word utterance. The red and green boxes indicate attended and unattended target words respectively}
\label{fig1_stimuli_illustration}
\end{figure}

An illustration of the stimuli for all paradigms is shown in \fref{fig1_stimuli_illustration}.

\subsection{Experimental setup and data acquisition}\label{ss_data_acquistition}
\begin{figure}[ht]
\centering
\includegraphics[width=1.0\linewidth]{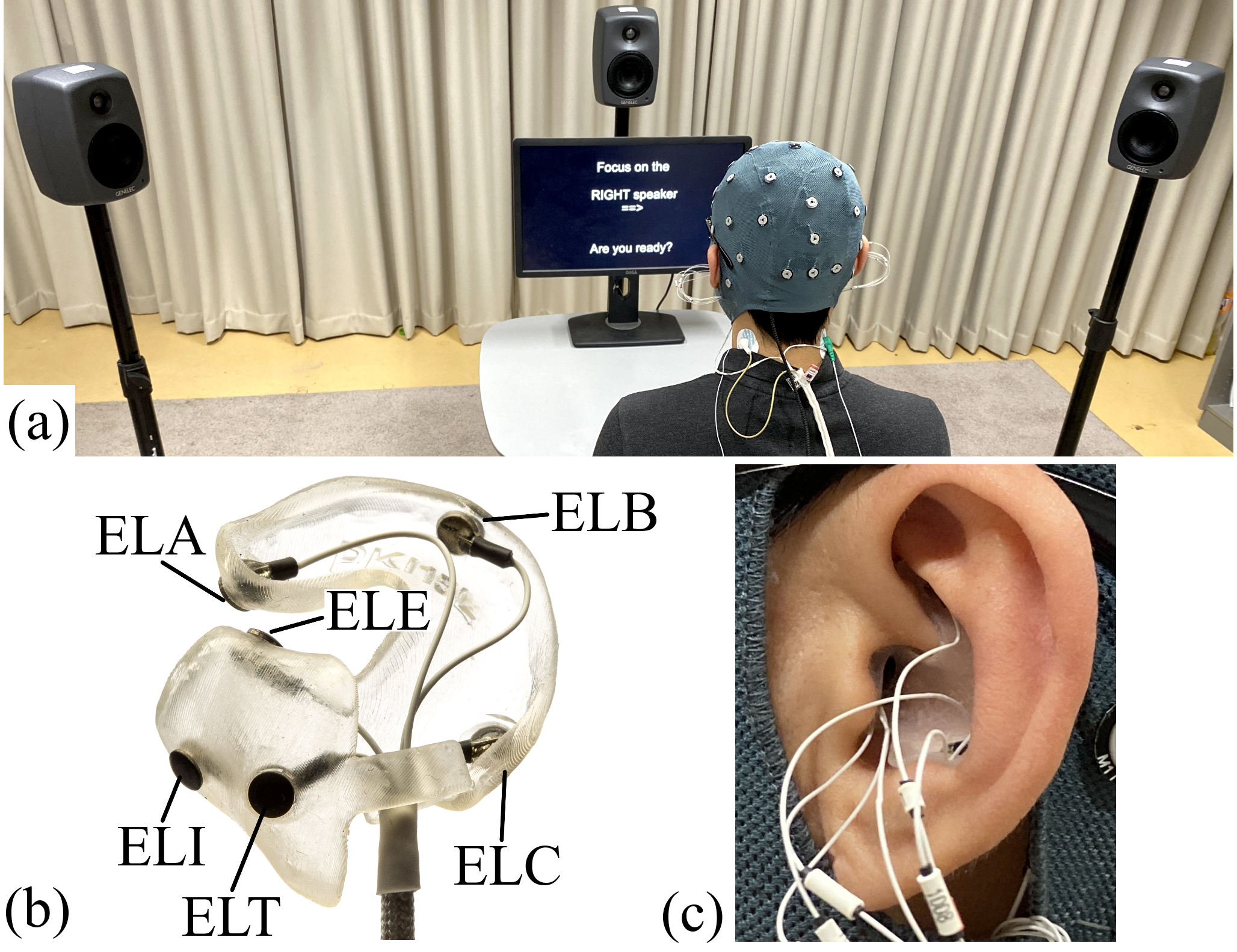}
\caption{Experimental setup: (a) Overall setup. (b) Earpiece for the left ear with electrodes inserted in positions A, B, C, T, E, and I. (c) An earpiece mounted in the ear.}
\label{fig2_experimental_setup}
\end{figure}

Participation in the study comprised two visits to the lab. At the first visit, an ear impression was taken to create individualized earpieces, which were used as part of the ear-EEG device \cite{kappelDryContactElectrodeEarEEG2019} to record EEG signals from the ears. The EEG was recorded at the second visit. The recording room was an acoustically shielded listening room with 0.4 seconds of reverberation time. The subject was seated in a chair in front of the loudspeakers, see \fref{fig2_experimental_setup}a. The EEG was recorded concurrently from 32 scalp electrodes and a left and right ear-EEG earpiece with six electrodes on each earpiece, see \fref{fig2_experimental_setup}b and \ref{fig2_experimental_setup}c. EEG data were collected using two TMSi Mobita amplifiers. Because the scalp and ear-EEG electrodes were made of different materials ($Ag$/$AgCl$ and $IrO_2$, respectively), they were connected to separate amplifiers. However, the two amplifiers shared a common electrode position that had a combined $Ag$/$AgCl$ and $IrO_2$ electrode, which was placed at the Fpz location. This allowed for the scalp and ear-EEG data to be combined during post-processing. The 32 scalp electrodes were named and located according to the 10/20 system. \Fref{fig2_experimental_setup} shows the details of the experimental setup. The paradigms were implemented in Python using the PsychoPy open-source platform v2022.1.4 \cite{peircePsychoPy2ExperimentsBehavior2019}.

The online sampling rate was set at 1000 Hz for both amplifiers and referenced to the average channels within each amplifier. The data were then off-line re-referenced to the common reference electrode, cut into trial blocks and concatenated to discard the data when subjects were in the relaxed state between trials. The data of each experiment were then saved to a different part in the final dataset under BIDS format \cite{gorgolewskiBrainImagingData2016}.

\subsection{Data analysis}\label{ss_data_analysis}
All twenty-four subjects completed all four experiments. The percentage of correct answers from each participant was used as a post hoc data exclusion criterion. In the following portion, we describe the common analysis pipeline utilized across all data analyses:
\begin{enumerate}
    \item Reference scalp EEG data to the averaged channel of all scalp channels and ear-EEG data to the averaged channel within each ear.\label{pp_1}
    \item Apply zero-phase FIR bandpass filter with passband corner frequencies 0.1 and 40 Hz.\label{pp_2}
    \item Apply the Independent Component Analysis (ICA) method to reject eye artifacts. (only for scalp EEG)\label{pp_3}
    \item Epoch data from -200 milliseconds to 1 second around the event onset and group into two experimental conditions. \label{pp_4}
    \item Apply baseline correction 200 milliseconds pre-stimulus.\label{pp_5}
    \item Apply peak-to-peak epoch rejection. Epochs containing min-max amplitude difference values above 200 $\mu$V were rejected.\label{pp_6}
    \item Calculate the individual average ERPs and grand average ERPs for each group by averaging across epochs within the group.
\end{enumerate}
In step (\ref{pp_3}), the ICA method was performed on the scalp EEG data using the {\it fastICA} method \cite{hyvarinenFastRobustFixedpoint1999}. ICs with a Pearson correlation coefficient with the EOG channel larger than 0.8 were considered as EOG artifacts. The measurement space signals were reconstructed, using the inverse weighting matrix, by leaving out the IC's identified as EOG artifacts. The processing pipeline was implemented using Python 3.9.9 and the supported library MNE-Python v1.2.0 \cite{gramfortMEGEEGData2013}. In step (\ref{pp_4}), the event onset was determined as the onset of the sound of the corresponding word in the speech by using the text-to-speech tool during the stimuli synthesis.

\subsection{Electrode configuration selection} \label{sss_electrode_selection}

\subsubsection{Scalp EEG.} \label{sss_scalp_config}
The initial analysis of scalp EEG was based on data recorded from the Pz electrode referenced to the average of the scalp electrodes. The choice of electrode configuration is justified by the design of the experimental paradigm resembling the structure of conventional P3b paradigms, from which it is well known that the peak of the P3b wave is largest in the central parietal region \cite{polichUpdatingP300Integrative2007}.

\subsubsection{Ear EEG.} \label{sss_ear_config}
In general, EEG signals and ERPs estimated from ear-EEG recordings have smaller amplitudes and lower signal-to-noise ratios as compared to scalp EEG. This is mainly due to small electrode distances, lower spatial coverage of the scalp, and higher electrode impedances \cite{kappelDryContactElectrodeEarEEG2019}. Additionally, due to inter-subject variability in the ear anatomy, there is a certain variation in the placement of the ear electrodes, which adds to the inter-subject variability of the ear-EEG signals. To optimize the SNR of the ERP waveform, we applied a spatial filter \cite{biesmansOptimalSpatialFiltering2015}. The spatial filter was optimized for each ear individually, and computed in the following way:

Let $C$ be the number of channels, $N$ be the number of samples within the duration of an epoch, $\mathcal{E}_\mathrm{T}$ and $\mathcal{E}_\mathrm{N}$ be the sets of the {\it{target}} and {\it{non-target}} epoch indexes, $n_\mathrm{T}$ and $n_\mathrm{N}$ be the number of the {\it{target}} and {\it{non-target}} epochs, $\bi{X}_i \in \mathbb{R}^{C\times N}$ be an epoch of multi-channels ear-EEG signals and $\bi{w} \in \mathbb{R}^{C\times 1}$ be the spatial filter (weighting vector). The goal of the spatial filtering method is to find the optimal $\bi{w}$ to maximize the ratio of output energy of the target ERP to the output of non-target ERP and other neural activities:

\begin{align}
\fl \bi{\hat{w}} &= \argmax_{\bi{w}} {\frac{\norm{\bi{X}_\mathrm{T}^\intercal\bi{w}}_2^{2}}{\frac{1}{n_\mathrm{N}}\sum\limits_{i \in \mathcal{E}_\mathrm{N}}{\norm{\bi{X}_i^\intercal\bi{w}}_2^{2}}}} \label{eq.1},
\end{align}\\
with $\bi{X}_\mathrm{T} = \frac{1}{n_\mathrm{T}}\sum\limits_{i \in \mathcal{E}_\mathrm{T}}{\bi{X}_i}, \bi{X}_\mathrm{T} \in \mathbb{R}^{C\times N}$ the average target epoch. \eref{eq.1} can be rewritten as:

\begin{align}
\fl \bi{\hat{w}} &= \argmax_{\bi{w}} {\frac{\bi{w}^\intercal\bi{X}_\mathrm{T}\bi{X}_\mathrm{T}^\intercal\bi{w}}{\frac{1}{n_\mathrm{N}}\sum\limits_{i \in \mathcal{E}_\mathrm{N}}{\bi{w}^\intercal\bi{X}_i\bi{X}_i^\intercal\bi{w}}}} \label{eq.2}\\
&= \argmax_{\bi{w}}\left( {n_\mathrm{N}\cdot\frac{\bi{w}^\intercal\bi{X}_\mathrm{T}\bi{X}_\mathrm{T}^\intercal\bi{w}}{\bi{w}^\intercal\bi{X}_\mathrm{N}\bi{X}_\mathrm{N}^\intercal\bi{w}}}\right) \label{eq.3}\\
&= \argmax_{\bi{w}}\left( {n_\mathrm{N}\cdot\frac{\bi{w}^\intercal\bi{R}_\mathrm{T}\bi{w}}{\bi{w}^\intercal\bi{R}_\mathrm{N}\bi{w}}} \label{eq.4}\right),
\end{align} \\
with $\bi{X}_\mathrm{N} \in \mathbb{R}^{C\times Nn_\mathrm{N}}$ the concatenation of non-target epochs along the time axes, $\bi{R}_\mathrm{T} \in \mathbb{R}^{C\times C}$ the auto-correlation matrix of averaged target epoch $\bi{X}_\mathrm{T}$ and $\bi{R}_\mathrm{N} \in \mathbb{R}^{C\times C}$ the auto-correlation matrix of $\bi{X}_\mathrm{N}$. From \eref{eq.4}, it appears that if $\bi{\hat{w}}$ is a solution, then any scalar multiplied with $\bi{\hat{w}}$ is also a solution. The unique solution can be obtained by introducing the constraint $\bi{w}^\intercal\bi{R}_\mathrm{N}\bi{w} = 1$ (i.e. keeping the output energy of the non-target ERPs constant). Thus, the optimal spatial filter can be found by solving the following constraint optimization problem:

\begin{align}
\fl \bi{\hat{w}} &= \argmax_{\bi{w}} {\bi{w}^\intercal\bi{R}_\mathrm{T}\bi{w}} \label{eq.5},
\end{align} \\
subject to the constraint\\
\begin{align}
\bi{w}^\intercal\bi{R}_\mathrm{N}\bi{w} = 1.\label{eq.6}
\end{align}

Solving the optimization problem described by \eref{eq.5} and \eref{eq.6} using the Lagrange multipliers method leads to a generalized eigenvalue problem with the solution being the eigenvector which corresponds to the largest generalized eigenvalue of $\bi{R}_\mathrm{T}$ and $\bi{R}_\mathrm{N}$.

As the spatial filter is optimized based on data, a rigorous cross-validation technique is necessary to avoid overfitting. Therefore, the ERP was estimated from the ear-EEG using the following cross-task validation scheme:

\begin{enumerate}
    \item  For each paradigm and for each ear (left and right), a spatial filter was estimated based on data from the other two paradigms. Each filter was trained using six channels within an ear referenced to the average channel of that ear.
    \item The spatial filters from step (i) were applied to all epochs from the paradigm that was not used for training those filters.
    \item The individual ERPs and the grand average ERPs were calculated from the output of step (ii) for each paradigm.
\end{enumerate}

\subsection{Statistical analysis}\label{ss_statistical_analysis}
Permutation and cluster-permutation tests \cite{marisNonparametricStatisticalTesting2007} were used for finding significant ERP responses. Specifically, to test the cognitive processing effect on the attended speech events at a particular time \textit{t}, the permutation test is performed in the following way:
\begin{enumerate}
    \item Calculate the test statistic (mean value) of the epoched data for each experimental condition (attended events and unattended events).
    \item Pool all epochs from both conditions.
    \item Randomly divide the epochs into two subsets, with the number of epochs in each subset equal to the number of epochs in each experimental condition.
    \item Calculate the test statistic (mean value) of each subset.
    \item Repeat steps (iii) and (iv) a large number of times (1000 times in this study) and construct the histogram of the test statistic.
    \item The p-value of the test is the proportion of the number of random partitions that resulted in data at least as extreme as the observed data from step (i).
\end{enumerate}

The cluster-based permutation test was used to determine the significance of the spatial and temporal locations of the effect, based on the ERP waveform and topography, respectively. The cluster-based permutation test follows a similar procedure to the permutation test described above. However, in the cluster-based permutation test, the test statistics are calculated at the cluster level. For instance, when testing the significance of the temporal location, the following steps are taken to find the test statistics for each permutation:
\begin{enumerate}
    \item For each sample in the epoch, calculate the difference between the mean values of two experimental conditions (attended events and unattended events).
    \item Find all clusters. A cluster is a set of temporally adjacent data points that have difference values larger (or absolutely larger for a two-sided test statistic) than a threshold.
    \item Calculate test statistics for each cluster by summing the mean difference within a cluster.
    \item The largest statistic among clusters is the permutation test statistic.
\end{enumerate}

\section{Results}\label{s_results}

\subsection{Paradigm 1: Word category oddball}\label{ss_p3s}

\begin{figure}[ht]
\centering
\includegraphics[width=\linewidth]{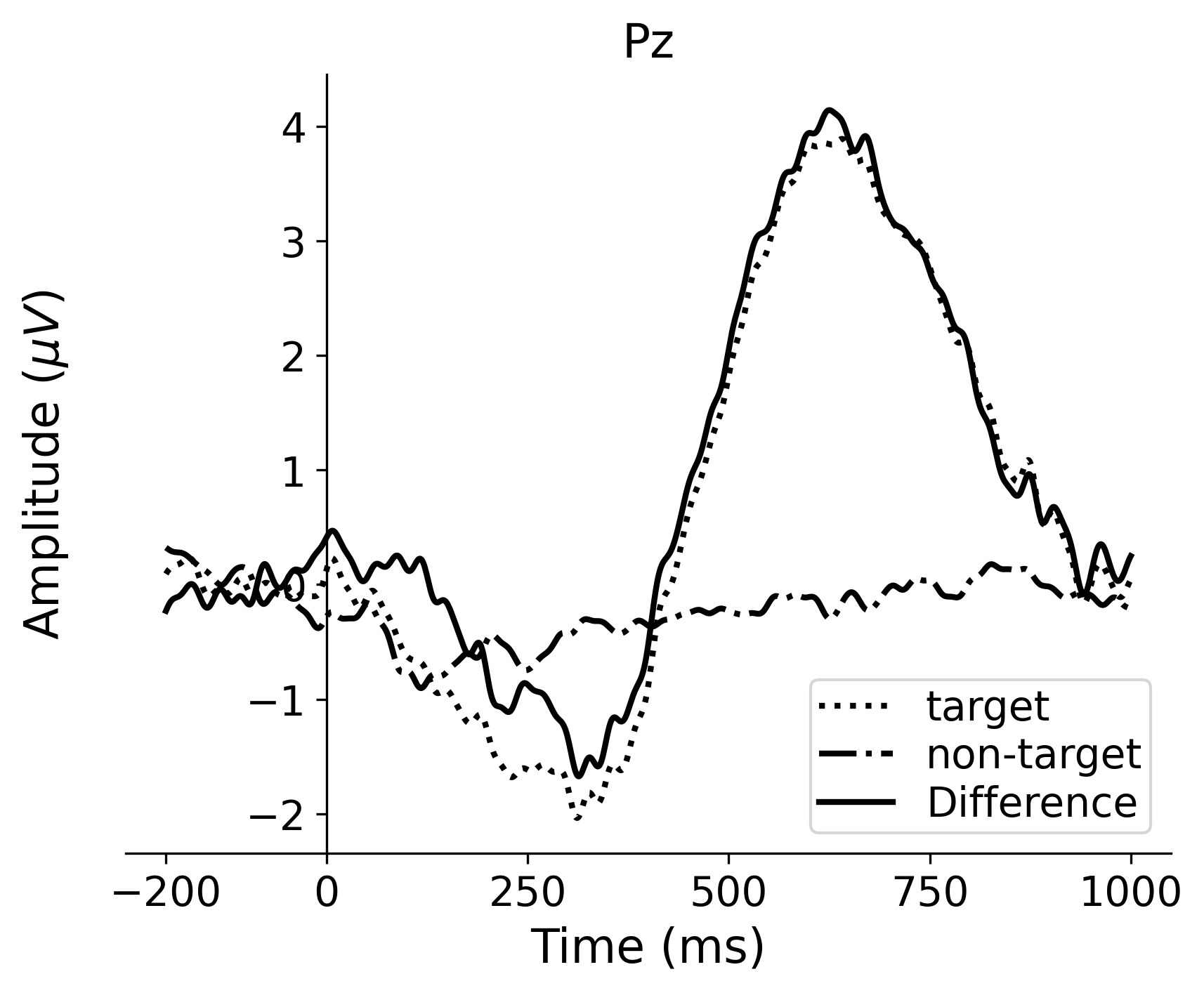}
\caption{Grand average ERP waveforms for paradigm 1. The ERPs are calculated from electrode Pz referenced to the average of all scalp electrodes.}
\label{fig3_P3s_grand_ERP_Pz}
\end{figure}

The individual ERPs were calculated by averaging across epochs corresponding to the target and non-target words, using the analysis pipeline described in \sref{ss_data_analysis}. The difference ERPs were obtained by subtracting the non-target ERPs from the target ERPs. The grand average ERPs of electrode Pz are shown in \fref{fig3_P3s_grand_ERP_Pz}. The grand average difference ERP peaks at 4.2 $\mu$V at around 620 ms, demonstrating a significant difference ($p<0.001$, permutation test for single time point) between the two experimental conditions. This positive going peak also appears consistently within the latency window 600-620 ms for all subjects (see \fref{fig_sup1_P3s_diff_ERP_Pz} in \ref{appd_A}).  To test the significance of the temporal location of this component, a cluster permutation test was performed as described in \sref{ss_statistical_analysis}. The threshold was selected as the maximum amplitude of the pre-stimulus grand average difference waveform to test all possible effects that appear after event onset. This threshold yields two clusters in the difference waveform, see \fref{fig5_p3s_pvalue}. The cluster permutation test results in a significant cluster within the time window 420-920 ms ($p = 0.001$), whereas the other cluster is not significant ($p = 0.158$). It should be noted that in \fref{fig5_p3s_pvalue}, only significant p-values are shown.

\begin{figure}[ht]
\centering
\includegraphics[width=1.0\linewidth]{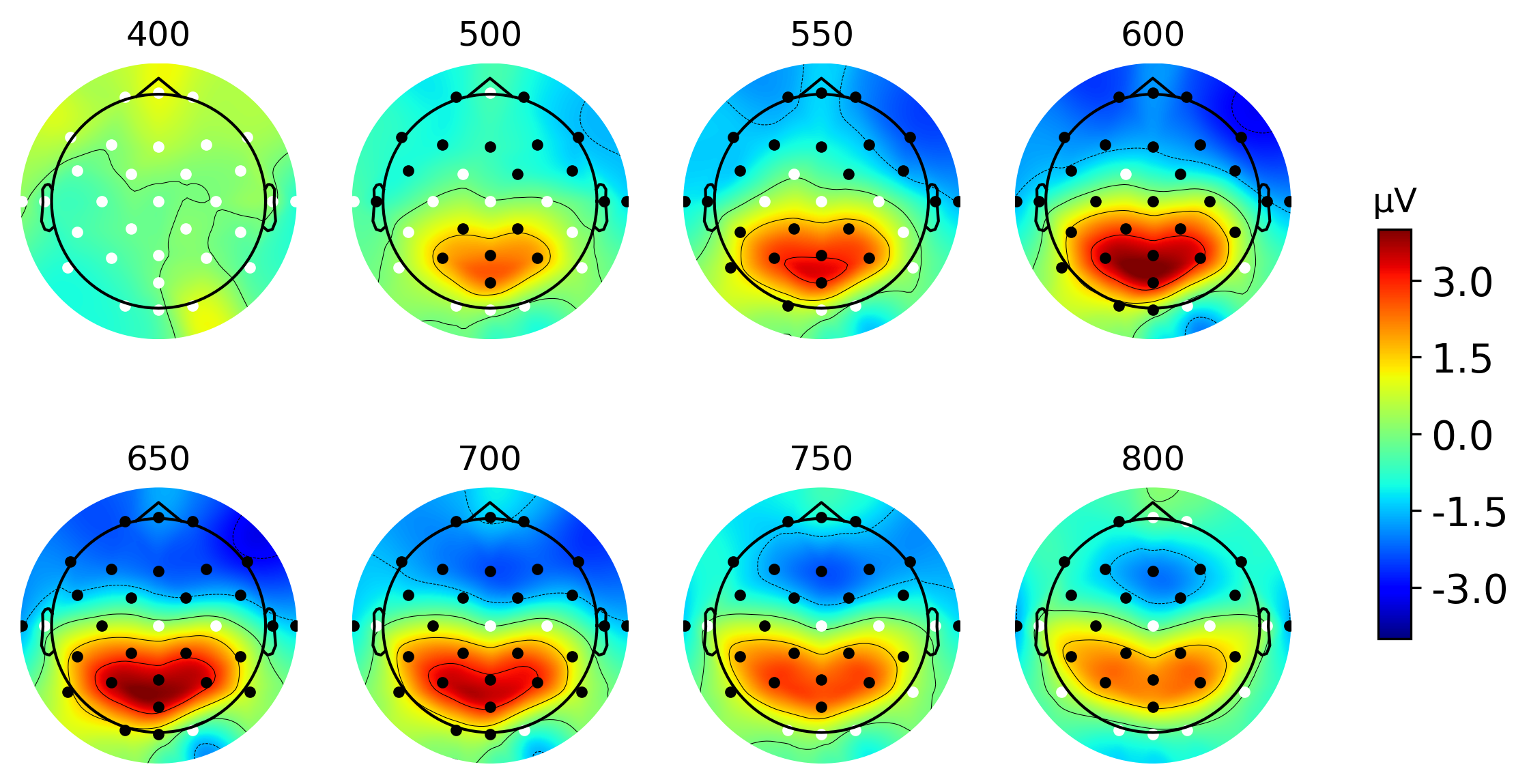}
\caption{Grand average scalp topographies from 400 to 800 ms post-stimulus of the difference ERP for paradigm 1. Black dots show significant electrodes and white dots show insignificant electrodes.}
\label{fig4_p3s_topo}
\end{figure}
\begin{figure}[ht]
\centering
\includegraphics[width=1.0\linewidth]{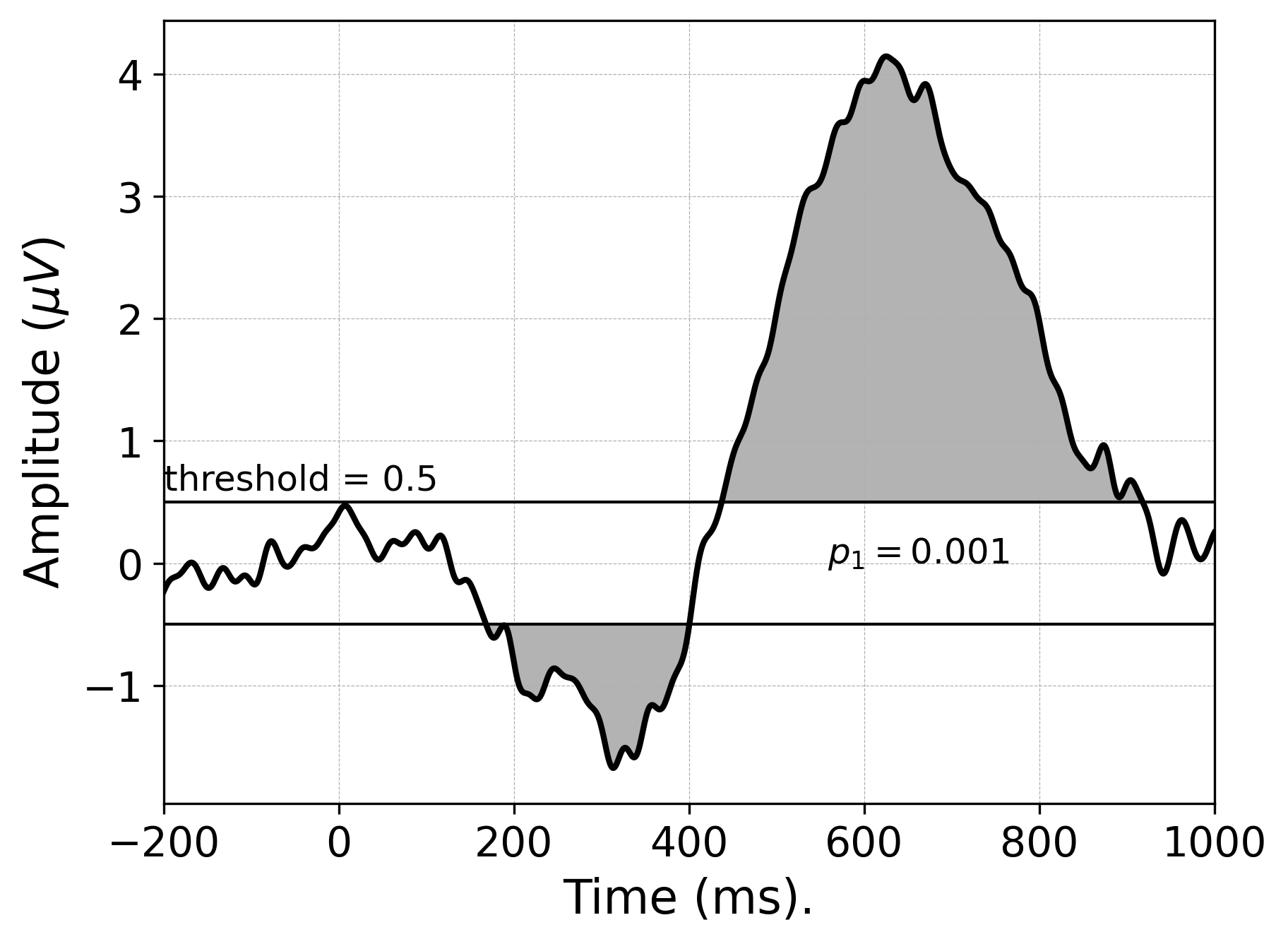}
\caption{P-values of permutation test to verify the significance of clusters across the time points for the difference ERP from electrode Pz referenced to the average of all scalp electrodes.}
\label{fig5_p3s_pvalue}
\end{figure}

\Fref{fig4_p3s_topo} represents the scalp topographies of the grand average difference ERP. The areas marked by black dots indicate significant differences in effects between target and non-target ERPs as determined by the spatial cluster permutation test described in \ref{ss_statistical_analysis}, using the same threshold selection procedure as mentioned above. We can see from \fref{fig4_p3s_topo} that the effect starts being significant from the timepoint 500 ms to 800 ms ($p<0.001$) at the parietal region. In each topography, there is also another significant cluster at the frontal region which likely represents the negative polarity of the effect.

\subsection{Paradigm 2 and 3: Competing streams} \label{sss_result_competing}
This section presents the results from the analysis of the data from experiment 2 and 3, the competing streams paradigms, using the analysis pipeline described in \sref{ss_data_analysis}.

\begin{figure}[ht]
\centering
\includegraphics[width=\linewidth]{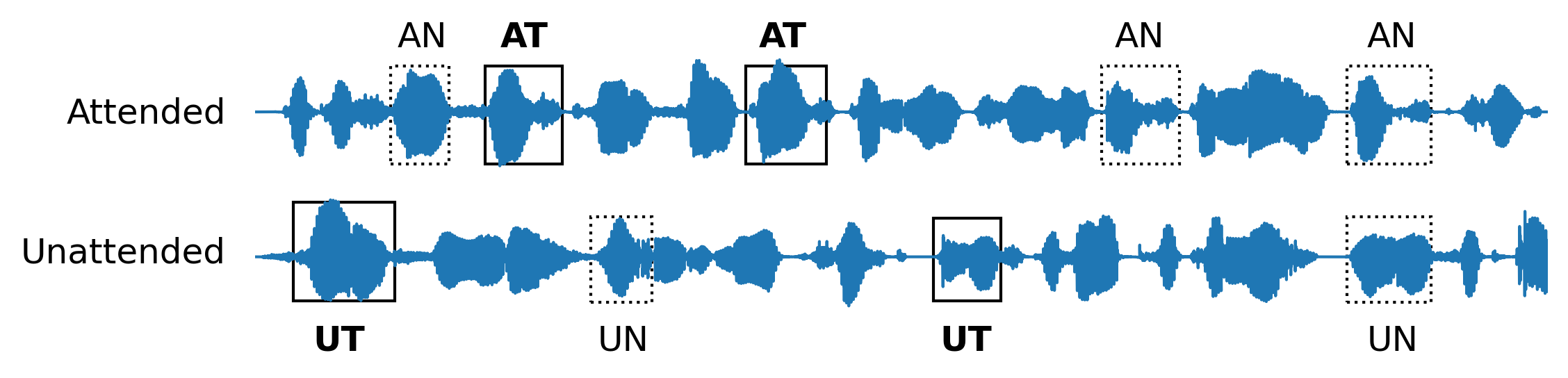}
\caption{Definition of different experimental conditions. AT: target event in the attended stream, AN: non-target event in the attended stream, UT: target event in the unattended stream, UN: non-target event in the unattended stream.}
\label{fig6_experiment_notation}
\end{figure}

To facilitate comparisons across different experimental conditions, notations AT (target event in the attended stream), AN (non-target event in the attended stream), UT (target event in the unattended stream), and UN (non-target event in the unattended stream) were used. \Fref{fig6_experiment_notation} shows an example of a stimulus used in experiment 3, which comprises continuous speech streams. The target events (AT, UT) were predefined as the specific word classes in each story with the remaining words in each stream considered as the non-target events (AN, UN). In experiment 2, all events were discrete spoken words. For the sake of clarity, in the remainder of this paper, the term {\it AT waveform} refers to the waveform calculated by averaging all epochs of {\it AT events}. The same rule applies to the AN, UT and UN waveforms. For better comparison, \fref{fig7_grand_ERP_comparisons_competing} shows the results of the analyses for both discrete and continuous speech paradigms.

\subsubsection{Paradigm 2: Word category with competing speakers} \label{sss_p3s_waveform_discrete_competing}
\begin{figure*}[ht]
 \centering
  \begin{subfigure}[ht]{\linewidth}
     \centering
     \includegraphics[width=\linewidth]{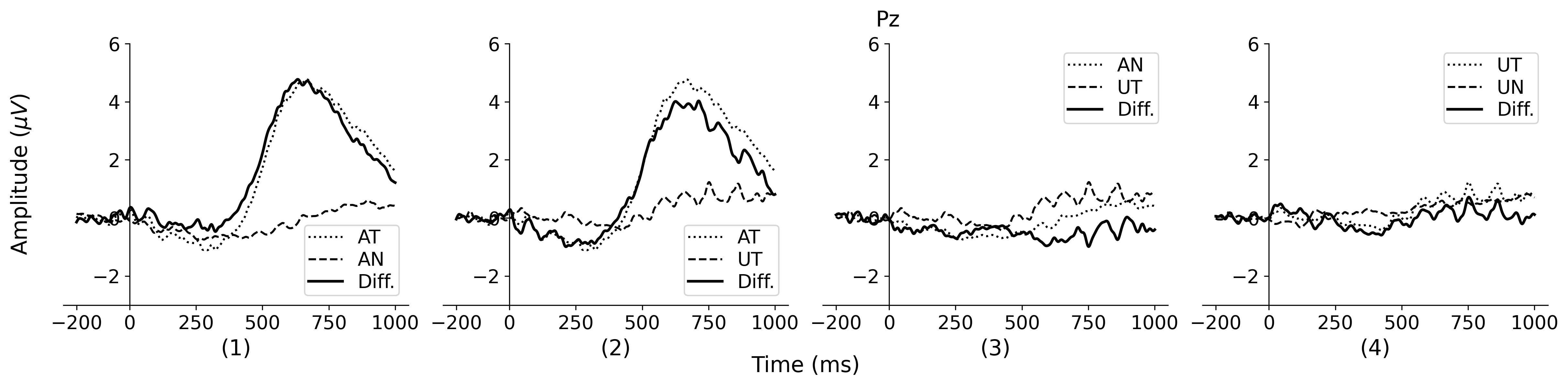}
     \caption{Discrete event competing stream paradigm.}
     \label{fig7a_discrete_comparisons}
 \end{subfigure}
 \hfill
 \begin{subfigure}[ht]{\linewidth}
     \centering
     \includegraphics[width=\linewidth]{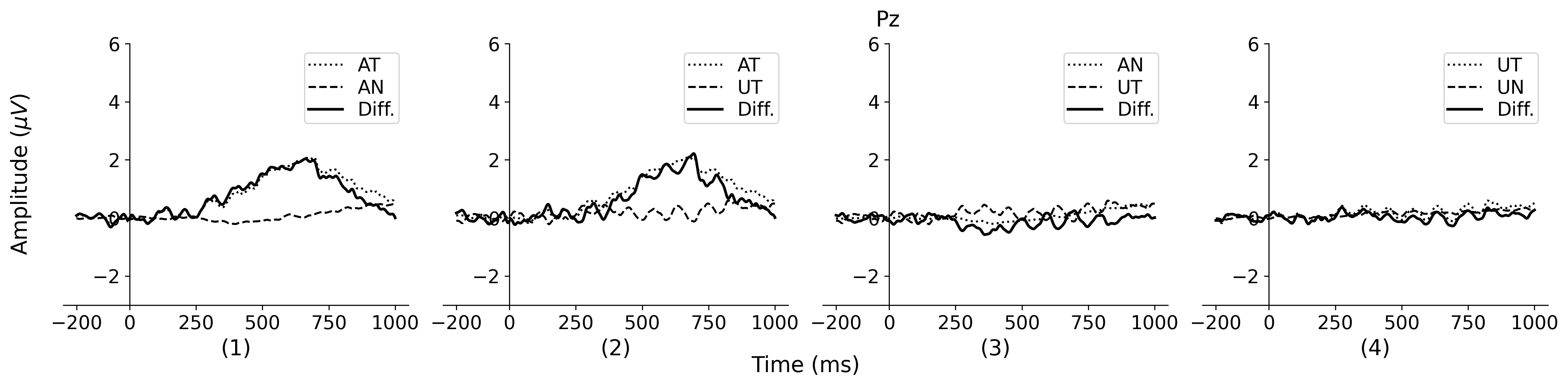}
     \caption{Continuous speech competing stream paradigm.}
     \label{fig7b_continuous_comparisons}
 \end{subfigure} 
\caption{The grand average ERPs of channel Pz in the multi-talker environment.}
\label{fig7_grand_ERP_comparisons_competing}
\end{figure*}

In experiment 2, the two competing streams consisted of sequences of discrete spoken words. Four different analyses were performed according to four pairs of experimental conditions AT vs. AN, AT vs. UT, AN vs. UT and UT vs. UN. The individual ERPs for each case were first calculated by following the processing pipeline described in \sref{ss_data_analysis}. The target and non-target ERPs corresponded to the former and the latter condition in each pair. The difference ERPs were obtained by subtracting the non-target ERPs from the target ERPs.  Finally, the grand average ERPs were calculated by averaging the individual ERPs. \Fref{fig7a_discrete_comparisons} shows the grand average ERPs for four analyses.

\Fref{fig7a_discrete_comparisons}.1 presents the results of the AT vs. AN comparison which compares the ERP waveforms of the target and non-target events in the attended stream. The target ERP (AT) has a clear positive deflection at around 620 ms, whereas there are no noticeable deflections in the non-target ERP (AN). The cluster permutation analysis of the difference ERP reveals a significant cluster in the 550 - 800 ms time window ($p < 0.005$).

Similarly, a comparison of the target ERPs in the attended stream (AT) and unattended stream (UT) are shown in \fref{fig7a_discrete_comparisons}.2, and the cluster permutation test reveals a significant difference in the time window 550-800 ms ($p < 0.001$).
The difference ERP between AN and UT is shown in \fref{fig7a_discrete_comparisons}.3; the cluster permutation test does not find a significant difference ($p>0.1$ for all time points). Finally, the difference between UT and UN is shown in \fref{fig7a_discrete_comparisons}.4, and as for the AN vs. UT, there is no significant difference ($p>0.1$ for all time points).

\begin{figure*}[ht]
 \centering
 \begin{subfigure}[ht]{0.495\linewidth}
     \centering
     \includegraphics[width=\linewidth]{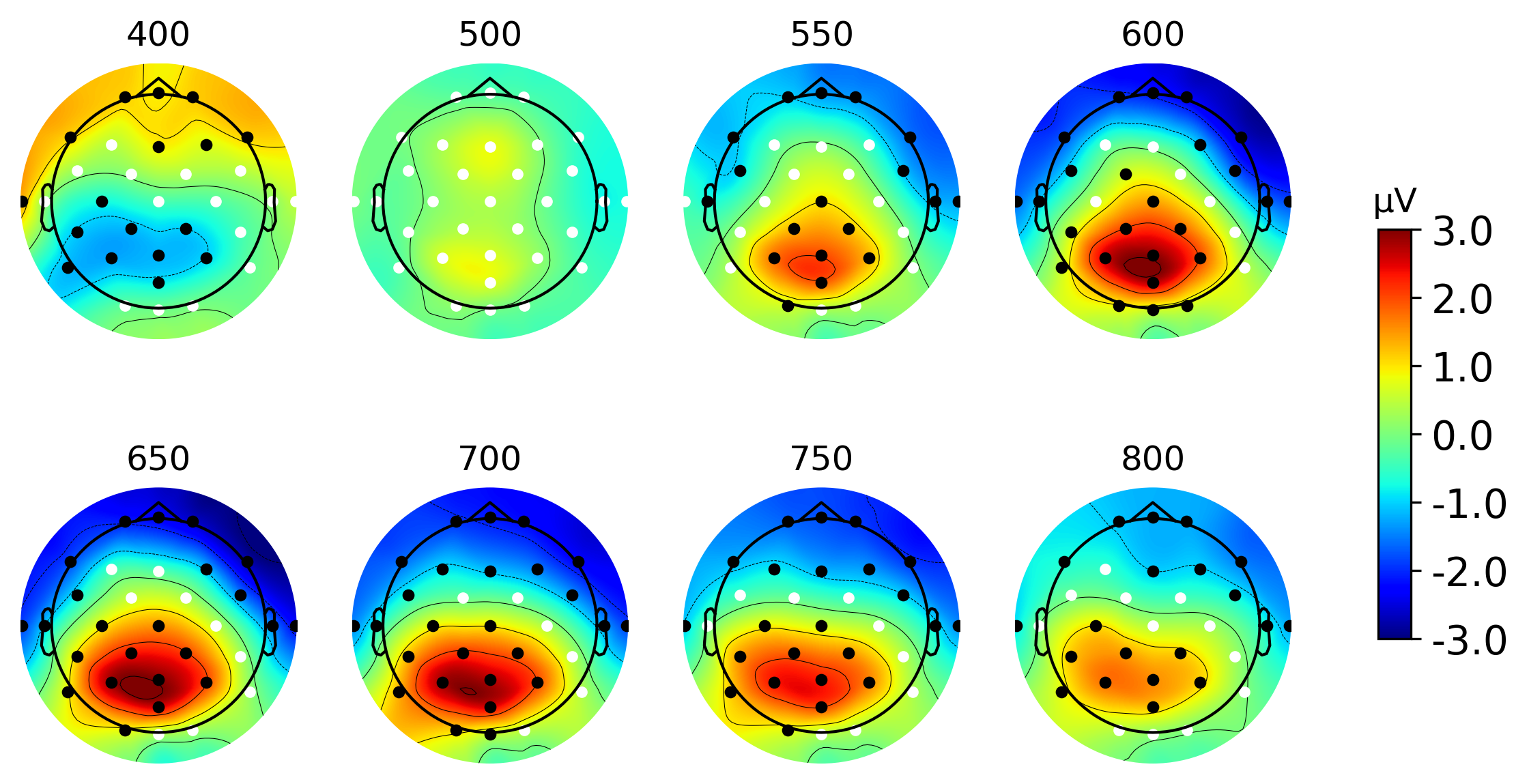}
     \caption{AT vs. AN}
     \label{fig8a_discrete_topo_AT_AN}
 \end{subfigure}
 \hfill
 \begin{subfigure}[ht]{0.495\linewidth}
     \centering
     \includegraphics[width=\linewidth]{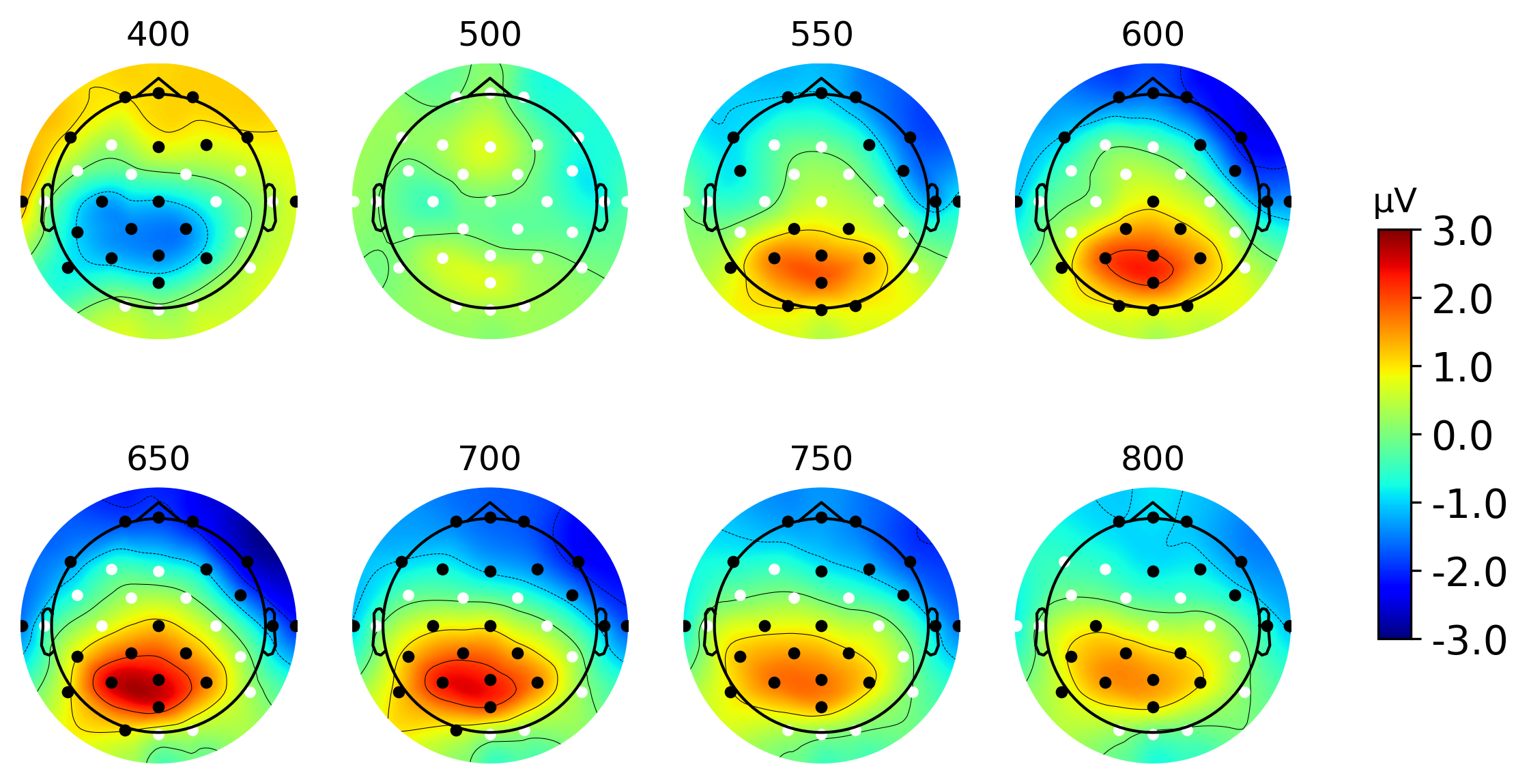}
     \caption{AT vs. UT}
     \label{fig8b_discrete_topo_AT_UT}
 \end{subfigure} 
\caption{Scalp topographies of AT-AN and AT-UT difference ERPs between 400 and 800 ms post-stimulus for discrete event competing paradigm. Black dots show significant electrodes, and white dots show insignificant electrodes.}
\label{fig8_topo_competing_discrete}
\end{figure*}

\Fref{fig8_topo_competing_discrete} shows the scalp topographies, between 400 and 800 ms, of the AT-AN and AT-UT difference ERPs. The permutation test shows that the significant clusters ($p < 0.01$ for the black dots region) of a cognitive component generated by AT events in the competing stream paradigm are also at the parietal site. This is completely consistent with the component in \sref{ss_p3s}.

It can be seen from the results of this paradigm that the cognitive ERP component can be elicited and observed at the parietal site even in a multi-talker environment if the brain performs a certain amount of attention to a specific event.

\subsubsection{Paradigm 3: Competing speech streams with targets} \label{sss_p3s_change_real_mechanism}
The data from experiment \ref{prdg_3}, which used the continuous speech competing paradigm, were used to perform the same four analyses as in \sref{sss_p3s_waveform_discrete_competing} using the same processing pipeline described earlier in \sref{ss_data_analysis}. The AN and UN waveforms were calculated based on the epochs of all the words that were not selected as the target events in the attended and unattended streams, respectively. This resulted in a significant increase in the number of AN and UN epochs.

\Fref{fig7b_continuous_comparisons} shows the grand average ERPs of four analyses for this paradigm. Similar to the results from the {\it word category with competing speakers} paradigm, the AT waveform is significantly different from the AN waveform ($p\leq 0.003$) and the UT waveform ($p\leq 0.01$) in the duration of 550-800 ms. However, the peak amplitude of the AT waveform in this paradigm is noticeably smaller (2.0 $\mu$V vs. 4.5 $\mu$V). The scalp topographies of two analyses: AT vs. AN and AT vs. UT show a similar pattern with that of the {\it word category with competing speakers} paradigm with the significant clusters at the parietal site (see \fref{fig_sup2_topo_competing_continuous} in \ref{appd_A}). There is no noticeable difference between the three types of events: AN, UT and UN.

Overall, we can see that attention to specific events can trigger the cognitive ERP component at the parietal site in the presence of competing speakers of both cases: discrete spoken words and continuous speech streams. In the competing speech streams paradigm, which is a more realistic listening situation, the amplitude of the component seems to be smaller.

\subsection{Cognitive component in ear-EEG}\label{ss_p3s_in_ear-EEG}
\begin{figure*}[ht]
\centering
\includegraphics[width=\linewidth]{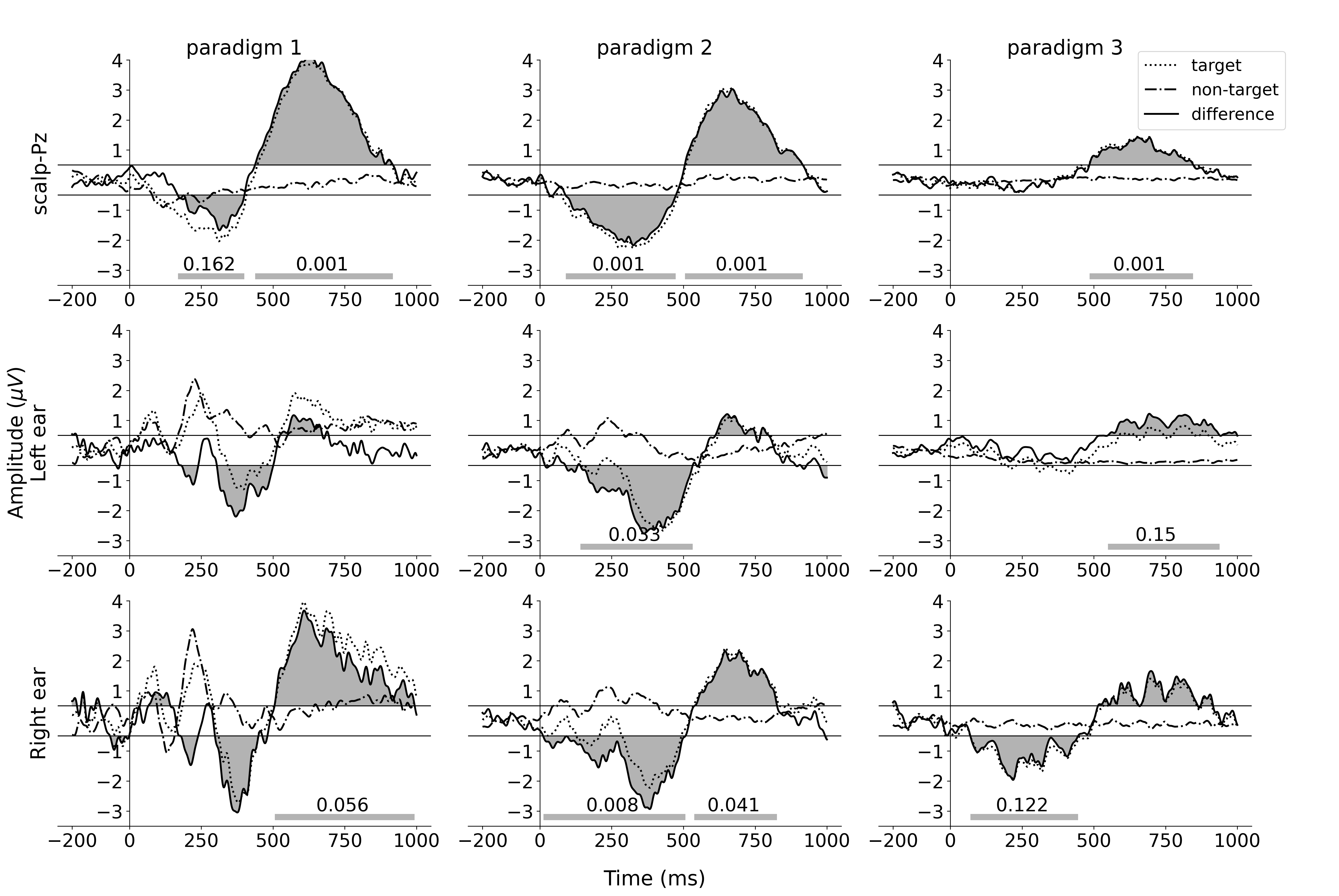}
\caption{ERP and difference waves for scalp and ear-EEG signals. The dotted lines represent the target ERP waveforms. The dot-dash lines represent the non-target ERP waveforms. The solid lines represent the difference ERP waveforms. The top row presents the grand average ERP waveforms calculated from electrode Pz, with reference to the average of all scalp electrodes. The middle row and bottom row present the grand average ERP waveforms obtained from the output of the spatial filter applied to 6 left-ear electrodes, with reference to the average of these electrodes, and the output of the spatial filter applied to 6 right-ear electrodes, with reference to these electrodes, respectively. The grey areas between the difference ERP waveforms and the horizontal lines $y = 0.5$/$y = -0.5$ represent the clusters detected by the 2-side permutation test with threshold 0.5. The cluster p-values lower than 0.2 are shown on the horizontal bars at the bottom of each subplot.}
\label{fig10_scalp-ear_comparison}
\end{figure*}

The ERP waveforms for the ear-EEG data were obtained using the spatial filtering method described in \sref{sss_ear_config}. Each column in \fref{fig10_scalp-ear_comparison} shows the grand average ERP waveforms for each paradigm. The ERP waveforms for electrode Pz are presented in the top row to facilitate comparison. The filtered grand average ERP waveforms for the left ear and the right ear are displayed in the middle row and bottom row, respectively.

While the cross-task validation scheme only finds one significant cluster (p = 0.041) from the right ear of the {\it word category with competing speakers} paradigm (paradigm 2), the difference ERP waveforms of both ears exhibit a very similar pattern to the Pz channel ERPs across all three paradigms. To quantify the similarity, we calculated the cosine similarity between the scalp (Pz) and the ear ERP’s, and the significance of the similarity was calculated from a permutation test distribution between the Pz and permuted target/non-target ear-ERPs. The result of this analysis is shown in table \ref{tabl1_similarity}. All ERP’s, except the left ear of the {\it word category oddball} paradigm (paradigm 1), are significant.

\begin{table}[]
\caption{\label{tabl1_similarity}Similarity scores and corresponding p-values of the ear and scalp ERPs for three paradigms.}
\begin{center}
\resizebox{\columnwidth}{!}{
\begin{tabular}{ |c|c|c|c|c|c|c| } 
 \hline
 \multirow{2}{*}{Ear} & \multicolumn{2}{c|}{paradigm 1} & \multicolumn{2}{c|}{paradigm 2} & \multicolumn{2}{c|}{paradigm 3} \\
 \cline{2-7} 
 \multirow{2}{*}{} & score & p-value & score & p-value & score & p-value \\ 
 \hline
 Left & 0.563 & 0.140 & 0.715 & 0.050 & 0.734 & 0.050 \\
 \hline
 Right & 0.802 & 0.017 & 0.910 & 0.001 & 0.787 & 0.022 \\ 
 \hline
\end{tabular}}
\end{center}
\end{table}

\section{Discussion}\label{discussion}
\subsection{Paradigm 1: Word category oddball}\label{ss_discuss_p3s}
By using discrete spoken word events and word categorization tasks in an oddball paradigm, a cognitive ERP component can be observed from the Pz electrode at around 620 ms in the results of this study. Although the peak appears with a significantly longer latency than reported in other studies (240-350 ms \cite{squiresTwoVarietiesLonglatency1975} and 200-500 ms \cite{polichUpdatingP300Integrative2007}), due to similarity in the experimental paradigm and the cognitive task evoking the response, we conjecture that the response and underlying mechanism is the same as for the P3b component. Moreover, a sequence of scalp topographies from 400 to 800 ms shown in \fref{fig4_p3s_topo} demonstrates that the positive deflection around 600 – 620 ms has a maximum in the central parietal region, which further supports and substantiates that the deflection reflects a P3b component. Therefore, in the context of this study, we name this component \textbf{\textit{P3s}} which refers to the late P3b generated by a cognitive task on speech events. According to Polich \cite{polichUpdatingP300Integrative2007}, the longer peak latency may be explained by higher stimulus evaluation timing due to the use of spoken word stimuli, task processing demand and higher task complexity. In this case, the task includes the semantic parsing of the meaning of the events and classifying them into two categories.

The presence of P3s component in the semantic oddball paradigm propels our comprehension of cognitive processing in response to speech events and opens up possibilities for employing ERP-based methodologies in speech-related cognitive research.

\subsection{Paradigm 2 and 3: Competing streams} \label{ss_discuss_competing}
This section discusses how the P3s component is impacted by both the multi-talker environment and normal speech context.

\subsubsection{Paradigm 2: Word category with competing speakers} \label{sss_discuss_competing_discrete}
The results shown in \fref{fig7a_discrete_comparisons} illustrate how the P3s component is affected by the attention in the multi-talker environment. The positive deflection of the difference ERP waveform of the AT and AN events shown in \fref{fig7a_discrete_comparisons}.1 suggests that the P3s component also can be elicited by the {\it word category oddball} in the presence of competing speakers. The unattended stream in this case plays a role as a distractor or background noise. The significant difference between AT and UT in the duration 550-800 ms ($p < 0.001$), see \fref{fig7a_discrete_comparisons}.2, suggests that the P3s component is specifically evoked by the attended oddballs, rather than by the presence of a semantic oddball itself. This is a remarkable characteristic making it a promising component for addressing the AAD problem. Moreover, we did not find any significant differences between the AN and the UT, which indicates that the target events in the unattended stream do not attract more attention than the non-target events in the attended stream. The result from \ref{fig7a_discrete_comparisons}.4 demonstrates that when the subject ignores the sound source, all the events in that source do not generate any difference in ERP waveforms and are almost equally unattended. This again clarifies the aforementioned hypothesis that the selective attention mechanism plays an important role in triggering and manipulating the P3s component rather than the presence of a semantic oddball itself.

\subsubsection{Paradigm 3: Competing speech streams with targets.} \label{sss_discuss_competing_continuous}
The results in \sref{sss_p3s_waveform_discrete_competing} show that identical P3s can be observed in a multi-talker environment. However, the stimuli, sequences of discrete speech events, were quite simple and well-formatted, creating an unreal listening context. Therefore, it is premature to conclude that the P3s component is a solid and promising candidate to apply in any speech-based BCI application. 

In \sref{sss_p3s_change_real_mechanism}, we presented an analysis of how the P3s manifests in a "more realistic" listening environment using natural continuous speech stimuli. It can be seen from \fref{fig7_grand_ERP_comparisons_competing} that all the ERP waveform patterns in the continuous speech paradigm (bottom row) are case-wise similar to the result in the discrete speech paradigm (top row). However, the peak amplitude of the AT waveform in this paradigm is noticeably smaller. It is reasonable because of the less distinct nature of the target events in the realistic context. Additionally, in the context of listening to a story, subjects tend to be less attended to the target events due to lack of time. The topographies again confirm that the neural source location when we use the natural speech stimuli is also at around the parietal site which is identical to the location of the P3s component in single stream paradigm (\fref{fig4_p3s_topo}) and in the discrete event competing streams (\fref{fig8_topo_competing_discrete}).

The results from this paradigm can be interpreted that the P3s component is observable even in a realistic listening context where the subject attends to one speech stream among multiple streams. The temporal and spatial location of P3s is independent of the complexity of the stimuli. Also, the level of attention significantly affects the amplitude of this component. In other words, the higher the attention allocated by the brain to the events, the larger the amplitude of the component. It is obvious that the target words in the attended streams, defined by the task in the experiment, become important words in the speech context. This can be further utilized to solve the AAD problem where we do not have predefined target words. It is very likely that in a specific speech context, the important words catch the most attention and evoke the larger P3s component. We hope to test this in future work.

\subsection{P3s in the ear-EEG signal}\label{ss_ear-EEG_discuss}
The findings from the analysis of the ear-EEG signals in section \ref{ss_p3s_in_ear-EEG} indicate a consistent presence of the P3s component across various cognitive tasks in the in-ear EEG recordings. Despite the inherent challenges of high variance and low SNR in the in-ear EEG data, the application of spatial filtering yielded similar in-ear ERPs to those observed on the scalp for all three paradigms ($p \leq 0.05$ except the left ear, paradigm 1). This demonstrates that the potential of the P3s component, elicited by speech events, is not only distributed spatially across scalp locations but also extends to both ears and can be consistently measured by in-ear electrodes and decoded by spatial filters.

Although the amplitude of the P3s component may be less distinct and not statistically significant in the in-ear EEG data, the investigation of this component opens up possibilities for utilizing a compact and miniaturized in-ear EEG setup for decoding auditory attention and facilitating other BCI applications.

Results from the statistical test show that the P3s component is significant in training tasks but not in validation tasks. This observation suggests a potential issue of overfitting the spatial filters due to the limited amount of training data. We believe that this problem could be mitigated by increasing the volume of training data. Furthermore, the low SNR characteristic of Ear-EEG could present a barrier to achieving significant results. This reveals challenges for future research to explore more advanced signal processing methods to utilize the P3s component in the in-ear EEG signals.

\section{Conclusion}\label{Conclusion}
Through the utilization of natural speech stimuli and cognitive tasks, a study of cognitive components related to auditory attention was conducted on both scalp and in-ear EEG devices. The findings demonstrate that the cognitive processing of natural speech events, specifically the P3s component, can be observed at parietal electrode sites and typically peaks at approximately 620 ms.  Furthermore, we have also shown that the P3s component is observable even in a multi-talker environment and its amplitude is influenced by the level of attention given by the brain to the speech events. These results suggest that the P3s component carries information that is useful for decoding auditory attention. Additionally, spatial filtering played a pivotal role in extracting the cognitive component from the ear-EEG signals, thus marking a significant advancement toward cognitively controlled hearing devices.

\datasttm
The dataset presented in this article is being used to address other research questions. Therefore, it is not publicly available upon publication. However, the data and the code for replicating the results can be provided upon request.

\ack
This work was funded by the William Demant Foundation, grant numbers 20-2673, and supported by Center for Ear-EEG, Department of Electrical and Computer Engineering, Aarhus University, Denmark.

\intrsttm
The author declares that the research was conducted without any conflict of interest.

\appendix
\section{Supplementary figures}\label{appd_A}
\begin{figure}[h!]
\centering
\includegraphics[width=\linewidth]{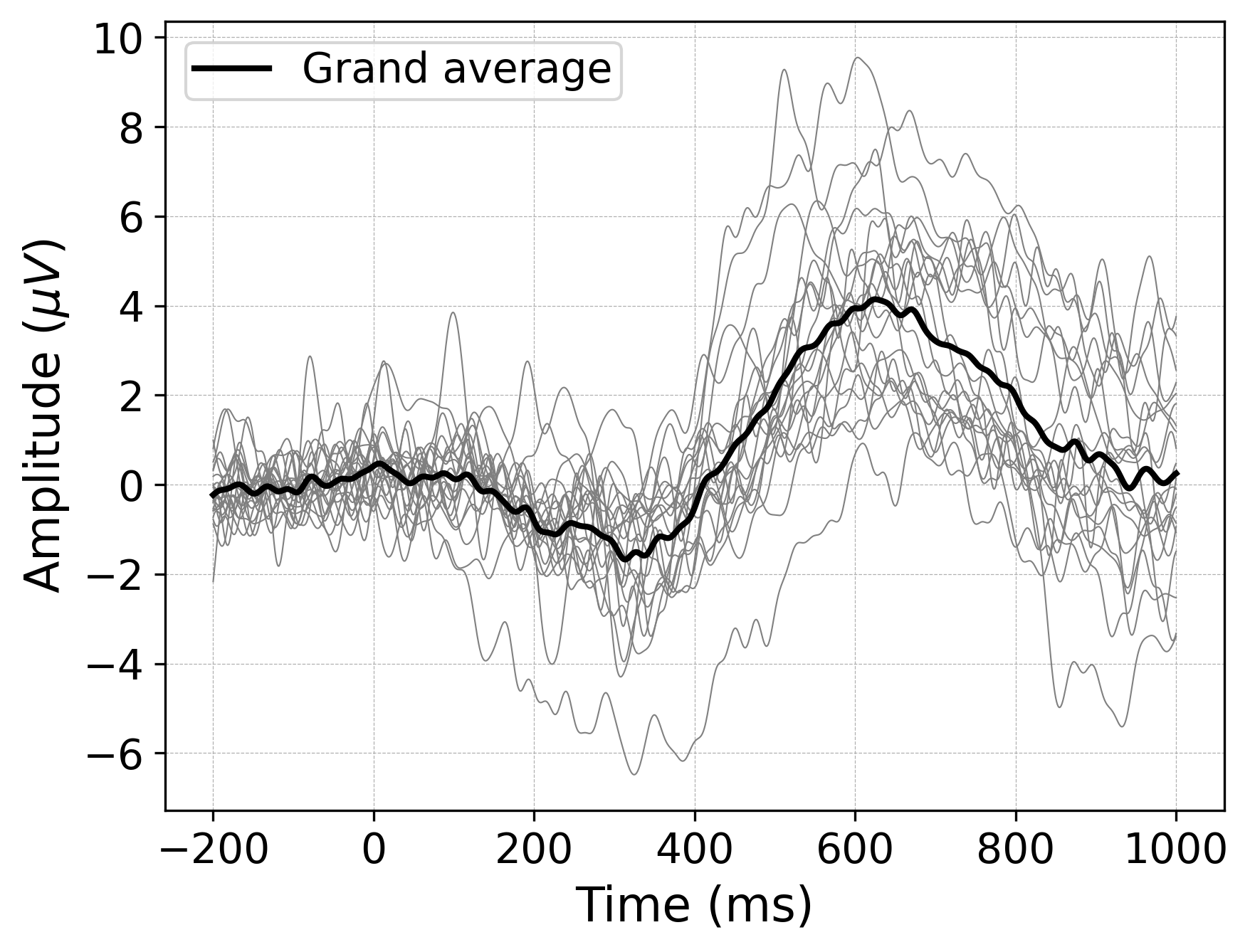}
\caption{Individual average and grand average ERP waveforms for the {\it word category oddball} paradigm. The ERPs are calculated from electrode Pz referenced to the average of all scalp electrodes.}
\label{fig_sup1_P3s_diff_ERP_Pz}
\end{figure}

\begin{figure*}[ht]
 \centering
  \begin{subfigure}[ht]{0.495\linewidth}
     \centering
     \includegraphics[width=\linewidth]{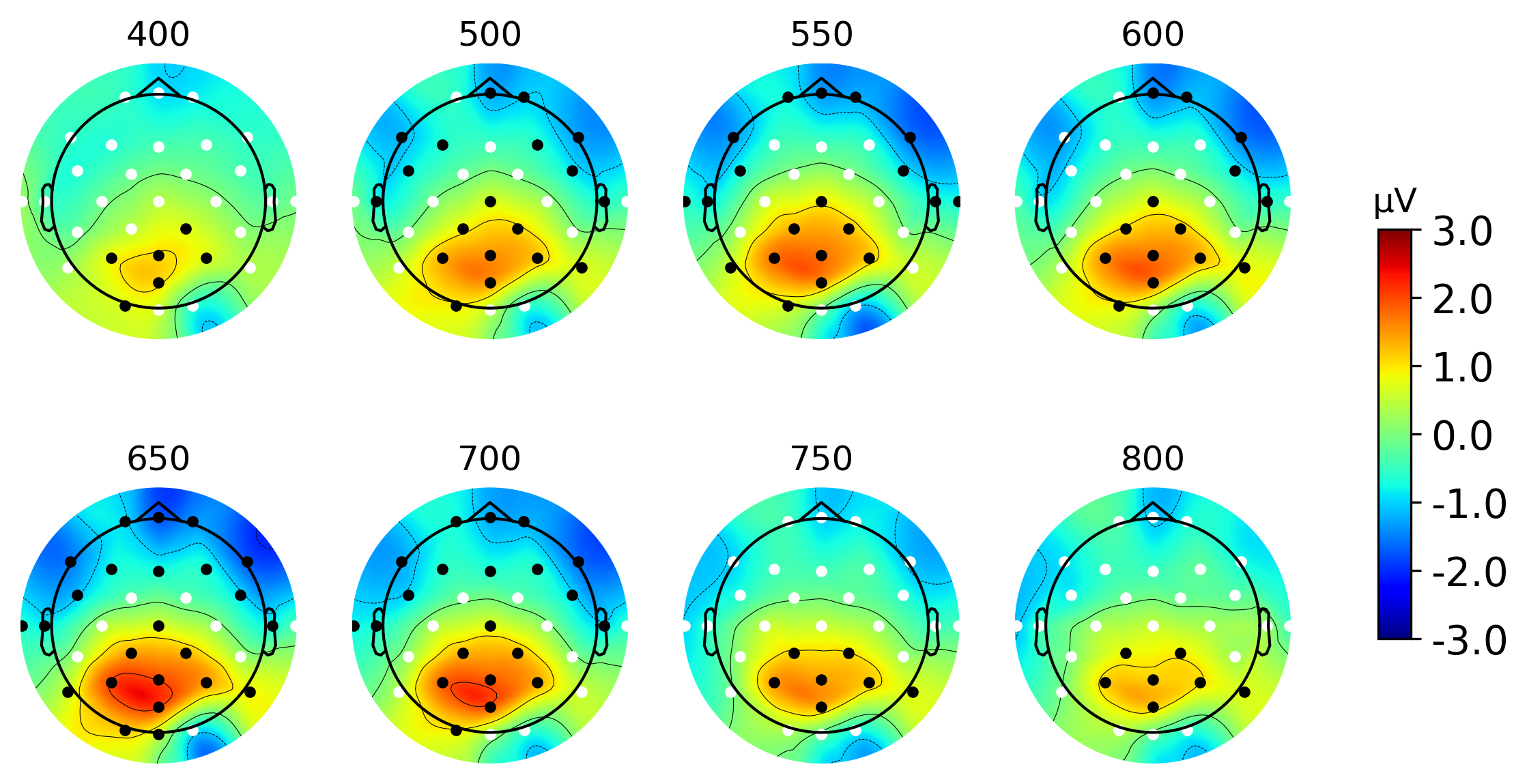}
     \caption{AT vs. AN}
     \label{fig_sup2a_speech_topo_AT_AN}
 \end{subfigure}
 \hfill
 \begin{subfigure}[ht]{0.495\linewidth}
     \centering
     \includegraphics[width=\linewidth]{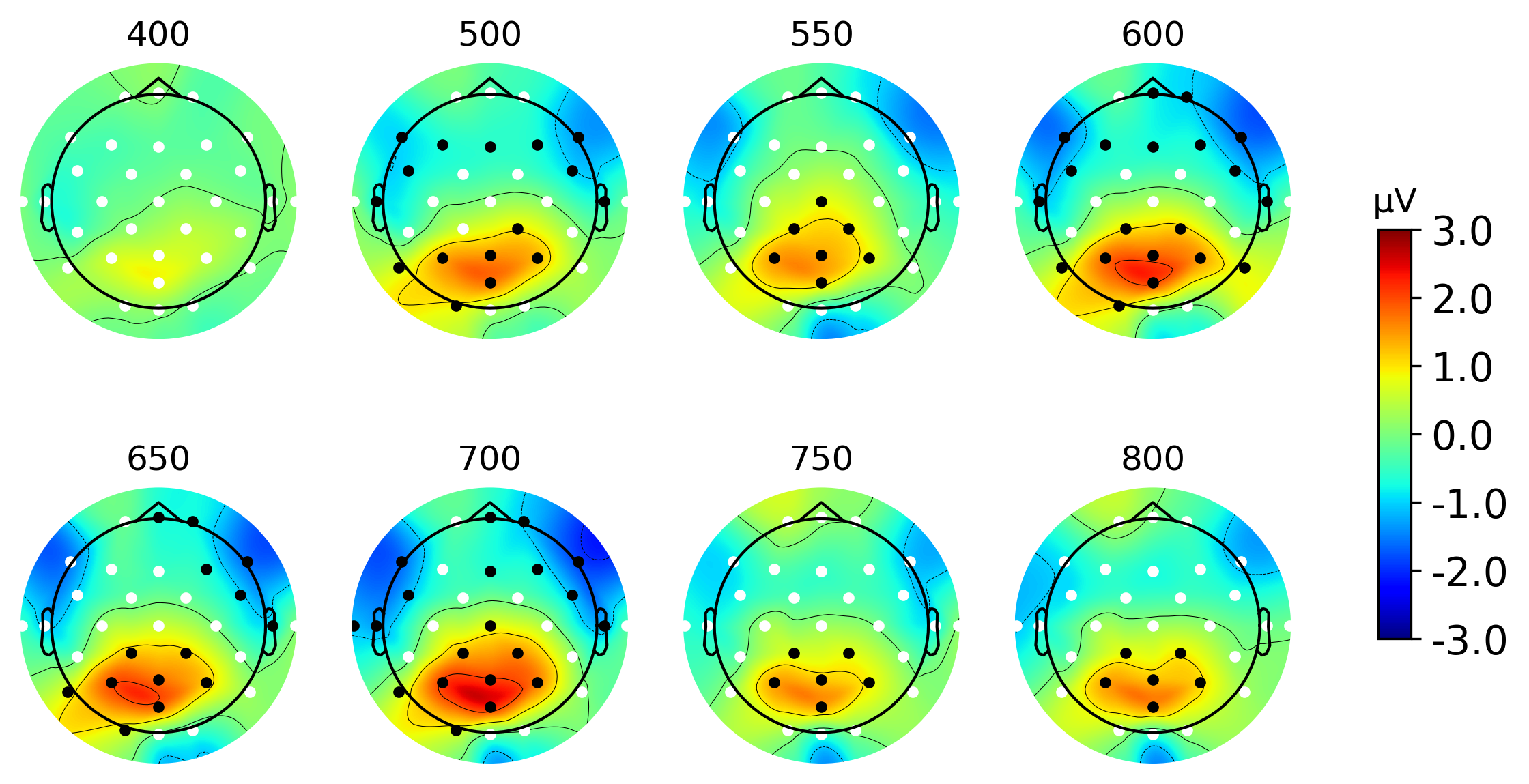}
     \caption{AT vs. UT}
     \label{fig_sup2b_speech_topo_AT_AN}
 \end{subfigure} 
\caption{Scalp topographies of AT-AN and AT-UT difference ERPs between 400 and 800 ms post-stimulus for competing speech streams paradigm. Black dots show significant electrodes, and white dots show insignificant electrodes.}
\label{fig_sup2_topo_competing_continuous}
\end{figure*}

\section*{References}
\bibliography{P3s_references}

\providecommand{\newblock}{}
\begin{thebibliography}{10}
\expandafter\ifx\csname url\endcsname\relax
  \def\url#1{{\tt #1}}\fi
\expandafter\ifx\csname urlprefix\endcsname\relax\def\urlprefix{URL }\fi
\providecommand{\eprint}[2][]{\url{#2}}

\bibitem{woodmanBriefIntroductionUse2010}
Woodman G~F 2010 A brief introduction to the use of event-related potentials in studies of perception and attention {\em Attention, Perception, \& Psychophysics\/} {\bf 72} 2031--2046 ISSN 1943-393X

\bibitem{broadbentPerceptionCommunication1958}
Broadbent D~E 1958 {\em Perception and Communication.\/} ({Elmsford}: {Pergamon Press})

\bibitem{treismanSELECTIVEATTENTIONMAN1964}
TREISMAN {\relax ANNE}~M 1964 {{SELECTIVE ATTENTION IN MAN}} {\em British Medical Bulletin\/} {\bf 20} 12--16 ISSN 0007-1420

\bibitem{treismanSelectiveAttentionPerception1967}
Treisman A and Geffen G 1967 Selective {{Attention}}: {{Perception}} or {{Response}}? {\em Quarterly Journal of Experimental Psychology\/} {\bf 19} 1--17 ISSN 0033-555X

\bibitem{hillyardElectricalSignsSelective1973}
Hillyard S~A, Hink R~F, Schwent V~L and Picton T~W 1973 Electrical {{Signs}} of {{Selective Attention}} in the {{Human Brain}} {\em Science\/} {\bf 182} 177--180

\bibitem{galambos40HzAuditoryPotential1981}
Galambos R, Makeig S and Talmachoff P~J 1981 A 40-{{Hz}} auditory potential recorded from the human scalp. {\em Proceedings of the National Academy of Sciences\/} {\bf 78} 2643--2647

\bibitem{aikenHumanCorticalResponses2008}
Aiken S~J and Picton T~W 2008 Human cortical responses to the speech envelope {\em Ear and Hearing\/} {\bf 29} 139--157 ISSN 0196-0202

\bibitem{dingEmergenceNeuralEncoding2012}
Ding N and Simon J~Z 2012 Emergence of neural encoding of auditory objects while listening to competing speakers {\em Proceedings of the National Academy of Sciences\/} {\bf 109} 11854--11859

\bibitem{osullivanAttentionalSelectionCocktail2015}
O'Sullivan J~A, Power A~J, Mesgarani N, Rajaram S, Foxe J~J, {Shinn-Cunningham} B~G, Slaney M, Shamma S~A and Lalor E~C 2015 Attentional {{Selection}} in a {{Cocktail Party Environment Can Be Decoded}} from {{Single-Trial EEG}} {\em Cerebral Cortex\/} {\bf 25} 1697--1706 ISSN 1460-2199, 1047-3211

\bibitem{biesmansAuditoryInspiredSpeechEnvelope2017}
Biesmans W, Das N, Francart T and Bertrand A 2017 Auditory-{{Inspired Speech Envelope Extraction Methods}} for {{Improved EEG-Based Auditory Attention Detection}} in a {{Cocktail Party Scenario}} {\em IEEE Transactions on Neural Systems and Rehabilitation Engineering\/} {\bf 25} 402--412 ISSN 1558-0210

\bibitem{geirnaertUnsupervisedSelfAdaptiveAuditory2021}
Geirnaert S, Francart T and Bertrand A 2021 Unsupervised {{Self-Adaptive Auditory Attention Decoding}} {\em IEEE Journal of Biomedical and Health Informatics\/} {\bf 25} 3955--3966 ISSN 2168-2208

\bibitem{aroudiImpactDifferentAcoustic2019}
Aroudi A, Mirkovic B, De~Vos M and Doclo S 2019 Impact of {{Different Acoustic Components}} on {{EEG-Based Auditory Attention Decoding}} in {{Noisy}} and {{Reverberant Conditions}} {\em IEEE Transactions on Neural Systems and Rehabilitation Engineering\/} {\bf 27} 652--663 ISSN 1558-0210

\bibitem{osullivanNeuralDecodingAttentional2017}
O'Sullivan J, Chen Z, Sheth S~A, McKhann G, Mehta A~D and Mesgarani N 2017 Neural decoding of attentional selection in multi-speaker environments without access to separated sources {\em 2017 39th {{Annual International Conference}} of the {{IEEE Engineering}} in {{Medicine}} and {{Biology Society}} ({{EMBC}})\/} pp 1644--1647 ISSN 1558-4615

\bibitem{detaillezMachineLearningDecoding2020}
{de Taillez} T, Kollmeier B and Meyer B~T 2020 Machine learning for decoding listeners' attention from electroencephalography evoked by continuous speech {\em The European Journal of Neuroscience\/} {\bf 51} 1234--1241 ISSN 1460-9568

\bibitem{nogueiraDecodingSelectiveAttention2019}
Nogueira W, Dolhopiatenko H, Schierholz I, B{\"u}chner A, Mirkovic B, Bleichner M and Debener S 2019 Decoding selective attention in normal hearing listeners and bilateral cochlear implant users with concealed ear {{EEG}} {\em Frontiers in Neuroscience\/} {\bf 13} ISSN 1662-4548

\bibitem{xuAuditoryAttentionDecoding2022}
Xu Z, Bai Y, Zhao R, Zheng Q, Ni G and Ming D 2022 Auditory attention decoding from {{EEG-based Mandarin}} speech envelope reconstruction {\em Hearing Research\/} {\bf 422} 108552 ISSN 0378-5955

\bibitem{hardleCanonicalCorrelationAnalysis2007}
 2007 Canonical {{Correlation Analysis}} {\em Applied {{Multivariate Statistical Analysis}}\/} ed H{\"a}rdle W and Simar L ({Berlin, Heidelberg}: {Springer}) pp 321--330 ISBN 978-3-540-72244-1

\bibitem{decheveigneDecodingAuditoryBrain2018}
{de Cheveign{\'e}} A, Wong D~D~E, Di~Liberto G~M, Hjortkj{\ae}r J, Slaney M and Lalor E 2018 Decoding the auditory brain with canonical component analysis {\em NeuroImage\/} {\bf 172} 206--216 ISSN 1053-8119

\bibitem{geirnaertElectroencephalographyBasedAuditoryAttention2021}
Geirnaert S, Vandecappelle S, Alickovic E, De~Cheveigne A, Lalor E, Meyer B~T, Miran S, Francart T and Bertrand A 2021 Electroencephalography-{{Based Auditory Attention Decoding}}: {{Toward Neurosteered Hearing Devices}} {\em IEEE Signal Processing Magazine\/} {\bf 38} 89--102 ISSN 1053-5888, 1558-0792

\bibitem{morayAttentionDichoticListening1959}
Moray N 1959 Attention in {{Dichotic Listening}}: {{Affective Cues}} and the {{Influence}} of {{Instructions}} {\em Quarterly Journal of Experimental Psychology\/} {\bf 11} 56--60 ISSN 0033-555X

\bibitem{deutschAttentionTheoreticalConsiderations1963}
Deutsch J~A and Deutsch D 1963 Attention: {{Some}} theoretical considerations {\em Psychological Review\/} {\bf 70} 80--90 ISSN 1939-1471

\bibitem{luckERPComponentsSelective2011}
Luck S~J and Kappenman E~S 2011 {{ERP Components}} and {{Selective Attention}} {\em The {{Oxford Handbook}} of {{Event-Related Potential Components}}\/} ed Kappenman E~S and Luck S~J ({Oxford University Press}) p~0 ISBN 978-0-19-537414-8

\bibitem{graingerWatchingWordGo2009}
Grainger J and Holcomb P~J 2009 Watching the {{Word Go}} by: {{On}} the {{Time-course}} of {{Component Processes}} in {{Visual Word Recognition}} {\em Language and linguistics compass\/} {\bf 3} 128--156 ISSN 1749-818X

\bibitem{erlbeckTaskInstructionsModulate2014}
Erlbeck H, K{\"u}bler A, Kotchoubey B and Veser S 2014 Task instructions modulate the attentional mode affecting the auditory {{MMN}} and the semantic {{N400}} {\em Frontiers in Human Neuroscience\/} {\bf 8} 654 ISSN 1662-5161

\bibitem{isrealP300TrackingDifficulty1980}
Isreal J~B, Chesney G~L, Wickens C~D and Donchin E 1980 P300 and tracking difficulty: {{Evidence}} for multiple resources in dual-task performance {\em Psychophysiology\/} {\bf 17} 259--273 ISSN 1469-8986

\bibitem{kramerAnalysisProcessingRequirements1983}
Kramer A~F, Wickens C~D and Donchin E 1983 An {{Analysis}} of the {{Processing Requirements}} of a {{Complex Perceptual-Motor Task}} {\em Human Factors\/} {\bf 25} 597--621 ISSN 0018-7208

\bibitem{mangunAllocationVisualAttention1990}
Mangun G and Hillyard S 1990 Allocation of visual attention to spatial locations: {{Tradeoff}} functions for event-related brain potentials and detection performance {\em Perception \& Psychophysics\/} {\bf 47} 532--550 ISSN 1532-5962

\bibitem{kappelDryContactElectrodeEarEEG2019}
Kappel S~L, Rank M~L, Toft H~O, Andersen M and Kidmose P 2019 Dry-{{Contact Electrode Ear-EEG}} {\em IEEE Transactions on Biomedical Engineering\/} {\bf 66} 150--158 ISSN 1558-2531

\bibitem{mikkelsenEEGRecordedEar2015}
Mikkelsen K~B, Kappel S~L, Mandic D~P and Kidmose P 2015 {{EEG Recorded}} from the {{Ear}}: {{Characterizing}} the {{Ear-EEG Method}} {\em Frontiers in Neuroscience\/} {\bf 9} ISSN 1662-453X

\bibitem{farooqEarEEGBasedVisual2015}
Farooq F, Looney D, Mandic D~P and Kidmose P 2015 {{EarEEG}} based visual {{P300 Brain-Computer Interface}} {\em 2015 7th {{International IEEE}}/{{EMBS Conference}} on {{Neural Engineering}} ({{NER}})\/} pp 98--101 ISSN 1948-3554

\bibitem{christensenChirpEvokedAuditorySteadyState2022}
Christensen C~B, Lunner T, Harte J~M, Rank M~L and Kidmose P 2022 Chirp-{{Evoked Auditory Steady-State Response}}: {{The Effect}} of {{Repetition Rate}} {\em IEEE Transactions on Biomedical Engineering\/} {\bf 69} 689--699 ISSN 1558-2531

\bibitem{mikkelsenAutomaticSleepStaging2017}
Mikkelsen K~B, Villadsen D~B, Otto M and Kidmose P 2017 Automatic sleep staging using ear-{{EEG}} {\em BioMedical Engineering OnLine\/} {\bf 16} 111 ISSN 1475-925X

\bibitem{mikkelsenAccurateWholenightSleep2019}
Mikkelsen K~B, Tabar Y~R, Kappel S~L, Christensen C~B, Toft H~O, Hemmsen M~C, Rank M~L, Otto M and Kidmose P 2019 Accurate whole-night sleep monitoring with dry-contact ear-{{EEG}} {\em Scientific Reports\/} {\bf 9} 16824 ISSN 2045-2322

\bibitem{fiedlerSinglechannelInearEEGDetects2017}
Fiedler L, W{\"o}stmann M, Graversen C, Brandmeyer A, Lunner T and Obleser J 2017 Single-channel in-ear-{{EEG}} detects the focus of auditory attention to concurrent tone streams and mixed speech {\em Journal of Neural Engineering\/} {\bf 14} 036020 ISSN 1741-2552

\bibitem{holtzeEarEEGMeasuresAuditory2022}
Holtze B, Rosenkranz M, Jaeger M, Debener S and Mirkovic B 2022 Ear-{{EEG Measures}} of {{Auditory Attention}} to {{Continuous Speech}} {\em Frontiers in Neuroscience\/} {\bf 16} ISSN 1662-453X

\bibitem{TexttoSpeechAILifelike}
Text-to-{{Speech AI}}: {{Lifelike Speech Synthesis}} https://cloud.google.com/text-to-speech

\bibitem{peircePsychoPy2ExperimentsBehavior2019}
Peirce J, Gray J~R, Simpson S, MacAskill M, H{\"o}chenberger R, Sogo H, Kastman E and Lindel{\o}v J~K 2019 {{PsychoPy2}}: {{Experiments}} in behavior made easy {\em Behavior Research Methods\/} {\bf 51} 195--203 ISSN 1554-3528

\bibitem{gorgolewskiBrainImagingData2016}
Gorgolewski K~J, Auer T, Calhoun V~D, Craddock R~C, Das S, Duff E~P, Flandin G, Ghosh S~S, Glatard T, Halchenko Y~O, Handwerker D~A, Hanke M, Keator D, Li X, Michael Z, Maumet C, Nichols B~N, Nichols T~E, Pellman J, Poline J~B, Rokem A, Schaefer G, Sochat V, Triplett W, Turner J~A, Varoquaux G and Poldrack R~A 2016 The brain imaging data structure, a format for organizing and describing outputs of neuroimaging experiments {\em Scientific Data\/} {\bf 3} 160044 ISSN 2052-4463

\bibitem{hyvarinenFastRobustFixedpoint1999}
Hyvarinen A 1999 Fast and robust fixed-point algorithms for independent component analysis {\em IEEE Transactions on Neural Networks\/} {\bf 10} 626--634 ISSN 1941-0093

\bibitem{gramfortMEGEEGData2013}
Gramfort A 2013 {{MEG}} and {{EEG}} data analysis with {{MNE-Python}} {\em Frontiers in Neuroscience\/} {\bf 7} ISSN 1662453X

\bibitem{polichUpdatingP300Integrative2007}
Polich J 2007 Updating {{P300}}: An integrative theory of {{P3a}} and {{P3b}} {\em Clinical Neurophysiology: Official Journal of the International Federation of Clinical Neurophysiology\/} {\bf 118} 2128--2148 ISSN 1388-2457

\bibitem{biesmansOptimalSpatialFiltering2015}
Biesmans W, Bertrand A, Wouters J and Moonen M 2015 Optimal spatial filtering for auditory steady-state response detection using high-density {{EEG}} {\em 2015 {{IEEE International Conference}} on {{Acoustics}}, {{Speech}} and {{Signal Processing}} ({{ICASSP}})\/} pp 857--861 ISSN 2379-190X

\bibitem{marisNonparametricStatisticalTesting2007}
Maris E and Oostenveld R 2007 Nonparametric statistical testing of {{EEG-}} and {{MEG-data}} {\em Journal of Neuroscience Methods\/} {\bf 164} 177--190 ISSN 0165-0270

\bibitem{squiresTwoVarietiesLonglatency1975}
Squires N~K, Squires K~C and Hillyard S~A 1975 Two varieties of long-latency positive waves evoked by unpredictable auditory stimuli in man {\em Electroencephalography and Clinical Neurophysiology\/} {\bf 38} 387--401 ISSN 0013-4694

\end{thebibliography}

\end{document}